\documentclass[fleqn,usenatbib]{mnras}

\usepackage{newtxtext,newtxmath}
\usepackage{showyourwork}

\usepackage[T1]{fontenc}

\DeclareRobustCommand{\VAN}[3]{#2}
\let\VANthebibliography\thebibliography
\def\thebibliography{\DeclareRobustCommand{\VAN}[3]{##3}\VANthebibliography}

\usepackage{graphicx}	
\usepackage{amsmath}	
\usepackage{chemformula}

\newcommand{\taubootisb}{$\tau$ Bo\"otis b}
\newcommand{\taubootisStar}{$\tau$ Bo\"otis}

\newcommand{\logfs}{$\log_{10}\mathrm{f}_s$}

\newcommand{\KpVsys}{$K_\mathrm{P} - V_\mathrm{sys}$}

\newcommand{\Kp}{$K_\mathrm{P}$}
\newcommand{\Vsys}{$V_\mathrm{sys}$}

\newcommand{\logWater}{$\log_{10}(\mathrm{H_2O})$}
\newcommand{\logCO}{$\log_{10}(\mathrm{CO})$}
\newcommand{\logMethane}{$\log_{10}(\mathrm{CH_4})$}
\newcommand{\logAmmonia}{$\log_{10}(\mathrm{NH_3})$}
\newcommand{\logCarbondiox}{$\log_{10}(\mathrm{CO_2})$}
\newcommand{\logHydrogencyn}{$\log_{10}(\mathrm{HCN})$}

\newcommand{\CCtologL}{CC-to-$\log L$}

\newcommand{\Msun}{\ensuremath{\mathrm{M_{\odot}}}}

\newcommand{\Mjup}{\ensuremath{\mathrm{M_{J}}}}
\newcommand{\Rjup}{\ensuremath{\mathrm{R_{J}}}}

\newcommand{\kms}{\ensuremath{\mathrm{km\,s^{-1}}}}

\newcommand{\sigmatell}{\ensuremath{\mathrm{\sigma_{tell}}}}
\newcommand{\FpFs}{\ensuremath{F_{P}/F_{S}}}
\newcommand{\TP}{\ensuremath{\mathrm{P-T}}}

\newcommand{\Npca}{\ensuremath{\mathrm{N_{PCA}}}}

\title[C/O and metallicity of \taubootisb{}]{The mystery of water in the atmosphere of \taubootisb{} continues: insights from revisiting archival CRIRES observations}

\author[Panwar et al.]{
Vatsal Panwar,$^{1,2}$\thanks{E-mail: vatsal.panwar@warwick.ac.uk}
Matteo Brogi,$^{3,4}$
Siddharth Gandhi$^{1,2}$
Heather Cegla$^{1,2}$\thanks{UKRI Future Leaders Fellow}
Marina Lafarga$^{1,2}$
\\
$^{1}$Department of Physics, University of Warwick, Coventry, UK, CV47AL\\
$^{2}$Center for Exoplanets and Habitability, University of Warwick, Coventry, UK, CV47AL\\
$^{3}$Dipartimento di Fisica, Universit\`a degli Studi di Torino, via P. Giuria 1, Turin, I-10125, Italy \\
$^{4}$Osservatorio Astrofisico di Torino, INAF, via Osservatorio 20, Pino Torinese, I-10025, Italy \\
}

\date{Accepted 2024 October 9. Received 2024 September 6; in original form 2024 April 2}

\pubyear{2023}

\begin{document}
\label{firstpage}
\pagerange{\pageref{firstpage}--\pageref{lastpage}}
\maketitle

\begin{abstract}
The chemical abundances of gas-giant exoplanet atmospheres hold clues to the formation and evolution pathways that sculpt the exoplanet population. Recent ground-based high-resolution spectroscopic observations of the non-transiting hot Jupiter \taubootisb{} from different instruments have resulted in a tension on the presence of water vapour in the planet's atmosphere, which impact the planet's inferred C/O and metallicity. To investigate this, we revisit the archival CRIRES observations of the planet's dayside in the wavelength range 2.28 to 2.33 $\mu$m. We reanalyse them using the latest methods for correcting stellar and telluric systematics, and free-chemistry Bayesian atmospheric retrieval. We find that a spurious detection of \ch{CH4} can arise from inadequate telluric correction. We confirm the detection of CO and constrain its abundance to be near solar $\log_{10}(\mathrm{CO})$ = --3.44$^{+1.63}_{-0.85}$ VMR.  We find a marginal evidence for H$_{2}$O with $\log_{10}(\mathrm{H_{2}O})$ = --5.13$^{+1.22}_{-6.37}$ VMR. This translates to super solar C/O (0.95$^{+0.06}_{-0.31}$), marginally sub-solar metallicity (--0.21 $^{+1.66}_{-0.87}$). Due to the relatively large uncertainty on H$_{2}$O abundance, we cannot confidently resolve the tension on the presence of H$_{2}$O and the super-solar atmospheric metallicity of \taubootisb{}. We recommend further observations of \taubootisb{} in the wavelength ranges simultaneously covering CO and $\mathrm{H_{2}O}$ to confirm the peculiar case of the planet's super-solar C/O and metallicity.        

\end{abstract}

\begin{keywords}
exoplanets; planets and satellites: atmospheres; stars: solar-type
\end{keywords}


\section{Introduction}
The atmospheres of exoplanets bear imprints of the physical and chemical processes governing planetary formation and evolution. Constraining fundamental atmospheric chemical abundance parameters, e.g. C/O ratio and metallicity, can provide insight into the effect of disk composition \citep{oberg_effects_2011, mordasini_imprint_2016}, processes delivering solids in the form of pebbles or planetesimals \citep{booth_chemical_2017}, and migration histories \citep{madhusudhan_toward_2014, khorshid_simab_2022}. This has been one of the primary motivations for spectroscopic observations of exoplanet atmospheres that have successfully detected and measured the abundances of numerous atomic and molecular species. One of the most prolific observational techniques proven to achieve this is High-Resolution Cross-Correlation Spectroscopy \citep[HRCCS,][]{snellen_orbital_2010}, which has been used in the last decade to characterize atmospheres of both transiting and non-transiting exoplanets. HRCCS involves using spectrographs (R = $\lambda$/$\Delta\lambda$ > 20000) to separate the Doppler shifted planetary signal from telluric and stellar lines. The orbital motion of a planet induces a Doppler shift of planetary atmospheric emission or transmission lines (at the order of several km/s for hot Jupiters) which can be isolated in velocity space by measuring the cross-correlation between the data and planetary model. 

HRCCS has been used in the near-infrared to detect Carbon and Oxygen bearing molecular species in the atmospheres of hot Jupiters \citep[e.g.][]{de_kok_detection_2013, birkby_detection_2013}, and in the optical to both detect \citep[e.g.][]{hoeijmakers_atomic_2018, pino_neutral_2020} and measure the longitudinal asymmetry \citep[e.g.][]{ehrenreich_nightside_2020, gandhi_spatially_2022, prinoth_time-resolved_2023} of ionized and neutral metals in the atmospheres of ultra-hot Jupiters. Recent work has also demonstrated the capability of HRCCS in measuring the thermal structure and absolute chemical abundances \citep{brogi_retrieving_2019, gibson_detection_2020} at precisions competitive with space-based spectroscopy, \citep{line_solar_2021, august_confirmation_2023} enabling the joint-retrieval of ground and space data \citep{smith_combined_2023}. 

\begin{figure}
\includegraphics[width=0.5\textwidth]{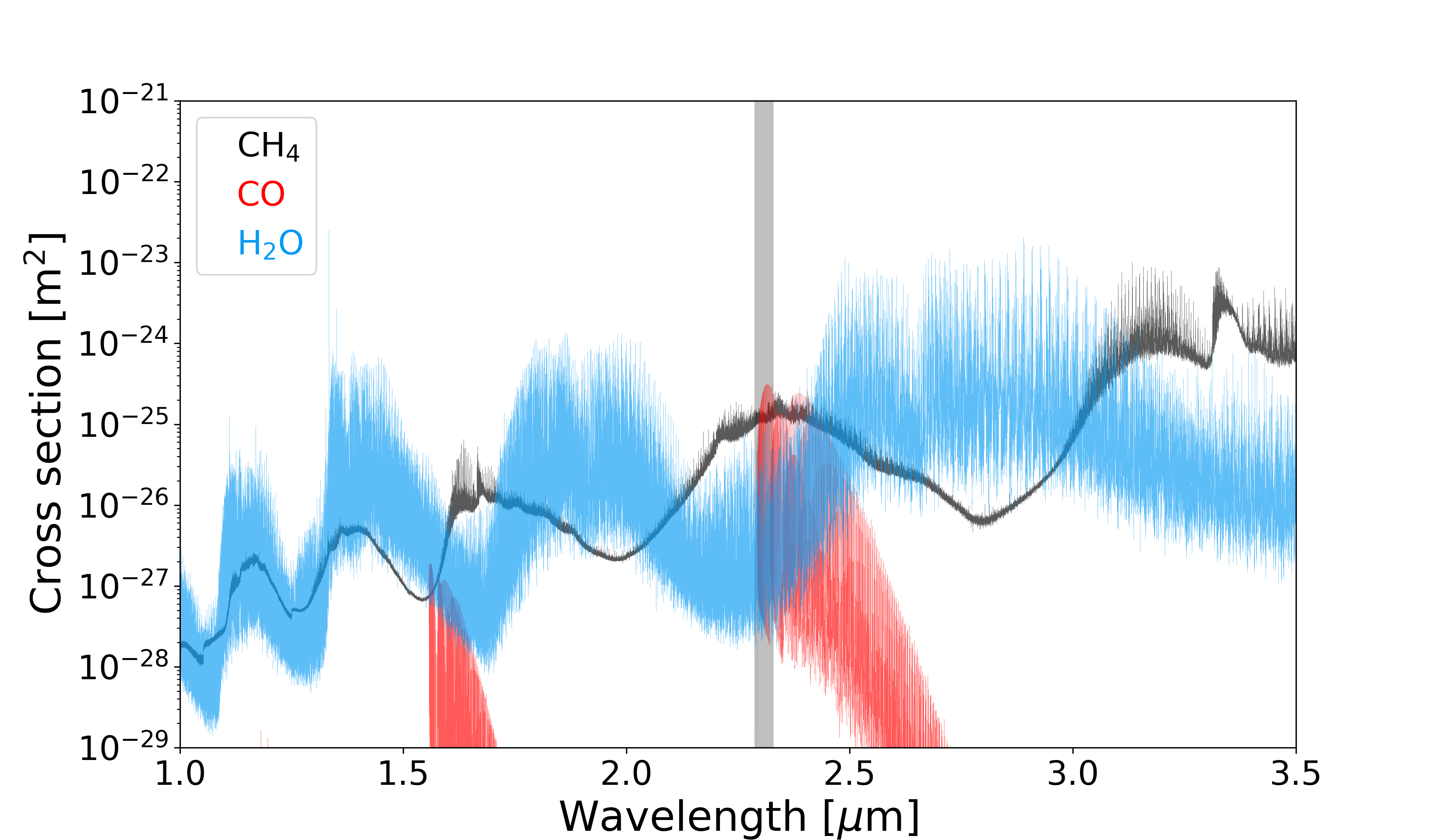}
\caption{Equilibrium chemistry abundance weighted cross-sections of \ch{H2O} (\logWater{} = -5), \ch{CH4} (\logMethane{} = -8), and CO (\logCO{} = -3.5) at 1 bar pressure and temperature 1600 K, in the 1 to 3.5 micron range at constant resolving power of 250000. The 2.28 to 2.33 $\mu$m range covered by the K band observations in this work (shaded grey region) is expected have large cross-sections from all the three species. The cross-sections have been computed from the latest line-list data available for each molecule as described in Section \ref{sec:model_calculation}.}
\label{fig:cross_sections}
\end{figure}

However, in some cases, discrepancies have emerged when observing the same target using multiple instruments. A stark example of this is the non-transiting hot-Jupiter \taubootisb{}. HRCCS is the only currently known method to characterize the atmosphere of a close-in non-transiting hot-Jupiter like \taubootisb{}. \taubootisb{} is one of the earliest hot-Jupiters detected using radial velocities around a nearby bright star (K$_{mag}$ = 3.5) \citep{butler_three_1997}, with an orbital inclination of 41.6$^{+1}_{-0.9}$ $^{\circ}$ and dayside equilibrium temperature T$_{eq}$ to be 1980 K or 1600 K assuming zero or perfect heat distribution to the night side respectively \citep{webb_water_2022}. The first atmospheric detection for \taubootisb{} was reported by \cite{brogi_signature_2012}, who used CRIRES observations to detect a strong absorption signal from atmospheric \ch{CO} emanating from the day side of the planet, consistent with a non-inverted temperature profile. These observations covered the K band where \ch{CO} is known to have large cross-section coming from the 2-0 R branch rotational-vibrational transitions of the \ch{CO} molecule starting around 2.29 $\mu$m as shown in Figure \ref{fig:cross_sections}. This wavelength region also contains relatively weaker cross-section from \ch{H2O} throughout, and in fact, the results from \cite{brogi_signature_2012} were inconclusive about the presence of signal due to \ch{water} in the data.   

Multiple studies after \cite{brogi_signature_2012} have searched for the signature of \ch{H2O} on the day side of \taubootisb{}. Targeting the 3 to 3.5 $\mu$m (L band) with Keck-NIRSPEC, \cite{lockwood_near-ir_2014} detected \ch{H2O} in the atmosphere of \taubootisb{}. In contrast, SPIRou observations by \cite{pelletier_where_2021} of the planet in the 0.9 to 2.5 $\mu$m range confirmed the detection of CO with super-solar abundance, but did not detect \ch{H2O}, and put a stringently low upper limit on the abundance of \ch{H2O}. \cite{pelletier_where_2021} concluded that \taubootisb{} has an atmosphere strongly depleted in \ch{H2O}, implying it has super-solar C/O and metallicity. More recently, observations using CARMENES by \cite{webb_water_2022} in the 0.9 to 1.7 $\mu$m detected signal from \ch{H2O} absorption.  

Detection or non-detection of water has strong implications on the planet's atmospheric metallicity and C/O, and hence on its inferred formation and evolution history. However, since detections of molecules cannot be directly compared to measurements of their abundances, solving the apparent discrepancy between different studies is not straightforward at face value. This motivated us to revisit the first high resolution spectroscopic dataset for \taubootisb{} published by \cite{brogi_signature_2012}, and reanalyse it using the now common principal component analysis (PCA) based detrending to remove stellar and telluric systematics. In addition to conducting a search for various molecular species, we also interpret the data through Bayesian atmospheric retrieval analysis based on the cross-correlation to log-likelihood mapping approach \citep{brogi_retrieving_2019}. Additionally, we use the latest data on molecular opacities available today, which are more accurate and complete than in 2012 when the first analysis for this data was done by \cite{brogi_signature_2012}. 

\section{Data Analysis}
\subsection{Observations}
\label{sec:observations}
We focus on observations obtained by program 186.C-0289(D) (PI: Snellen) and first published by \cite{brogi_signature_2012}. \taubootisStar{} was observed using CRIRES (along with the MACAO system) at VLT UT1 for three nights on 01-04-2011 to 02-04-2011, 08-04-2011 to 09-04-2011, and 14-04-2011 to 15-04-2011. Each night of observations lasted approximately 6 hours and covering the planetary orbital phase range of 0.37 to 0.63, and spanning the wavelength range of 2287 to 2345 nm over four CRIRES detectors. Exposure time per exposure was fixed to 10 seconds for the observations taken on night of 01-04-2011 and 14-04-2011, and 5 seconds for the night of 08-04-2011. The total number of stellar spectra obtained were 180, 80, and 193 for the dates 01-04-2011, 08-04-2011, and 14-04-2011 respectively. The observations were taken using a 0.2'' slit, yielding a resolution of R $\sim 100,000$. The telescope was nodded by 10'' between each frame in ABBA pattern to provide accurate sky background subtraction. 

As in the original analysis by \cite{brogi_signature_2012}, we reduced the using the CRIRES pipeline version v. 2.1.1, which performed the necessary steps of dark subtraction, flat-fielding, non-linearity corrections and bad-pixel removal, extraction of the 1D spectra. We obtain average signal-to-noise per exposure across the three nights of observations in the range of 130 to 250. 


Time dependent shifts in the wavelength solution were corrected for by aligning the spectra from all the exposures to the spectrum with the highest signal-to-noise.
The wavelength solution was then determined by fitting a model telluric spectrum computed via the ESO SkyCalc model calculator \citep{noll_atmospheric_2012, jones_advanced_2013} to the position of telluric lines in the data. All these steps are identical to \cite{brogi_signature_2012} and we refer the reader to the original publication for further details. As noted by \cite{brogi_signature_2012}, detector $\#$1 and detector $\#$4 in old CRIRES suffer from an odd-even effect because of the readout pattern being along the cross-dispersion direction for these detectors \footnote{\href{https://www.eso.org/sci/facilities/paranal/instruments/crires/doc/VLT-MAN-ESO-14500-3486_v87.pdf}{CRIRES Manual Period 87}}. This effect is smaller for exposure times $>$2 seconds and total signal $<$ 10000 ADU ($\sim$32000 e$^{-}$) which is the case for detector $\#$1 but not detector $\#4$ in our observations. Moreover, the majority of signal from \ch{CO} and \ch{H2O} covers the wavelength ranges spanned by the first three detectors. Hence, we omit the data from detector $\#4$ to avoid contribution from the odd-even effect and consider only the data from detectors $\#1$, $\#2$, and $\#3$.   

\subsection{Removing stellar and telluric systematics using PCA}
\label{sec:pca_detrending}
The dominant source of systematics in ground-based time series high resolution spectroscopic data is telluric contamination. In the wavelength range 2287 to 2330 nm (detector \#1, \#2, and \#3) probed by our observations, the primary telluric contamination is from absorption lines of \ch{H2O} and \ch{CH4} \citep{noll_atmospheric_2012, smette_molecfit_2015}. Changing precipitable water vapour (PWV) in the line of sight of observation induces variations in the depth of telluric lines of \ch{H2O} covered by the observations. These, along with the stellar lines and other instrumental variations in time, need to be corrected for before conducting a search for lines of \ch{H2O} originating from the planetary atmosphere. 

We use a principal component analysis (PCA) approach to remove the telluric and stellar contamination in time series spectral data cubes for each date. This approach was first used for such observations by \cite{de_kok_detection_2013}, and has become one of the most popular methods of choice for this purpose \citep[e.g.][]{giacobbe_five_2021, lafarga_hot_2023}. The main reason this approach works is that stellar and telluric lines are relatively stationary in wavelength dimension, whereas planetary lines for fast moving short period planets (orbital period $\leq$ 10 days) like hot Jupiters are subject to a Doppler shift that varies by tens of \kms{} during a course of a night of observation. PCA, which at the core uses a linear algebra algorithm called Singular Value Decomposition (SVD), computes a set of orthogonal basis vectors also called as eigenvectors or PCA components (each the same length as the number of exposures) for a data cube matrix per detector (with dimensions time $\times$ wavelength pixels). The idea is that a linear combination of these PCA components should form a close match to the contribution from the telluric and stellar lines in the data cube. PCA components can be ranked in order of the amount of variance of flux an eigenvector represents, which in our case is the variance of flux in the time dimension. We note that when PCA decomposes flux as a function of time (time being the independent variable) it is said to work in the time domain, which is indeed the case in this context. Vice versa, when wavelength is considered the independent variable, PCA works in the wavelength domain.

Our implementation of PCA in this work has two additional steps to SVD \citep{de_kok_detection_2013} - standardization, and multi-linear regression. We summarize them both here.

Working on each detector and each date independently, we perform the first step to `standardize' the reduced flux data cubes obtained in Section \ref{sec:observations} by subtracting the respective mean flux from the time series for each spectral channel and then dividing this difference by the standard deviation of the flux in the channel. This step of standardization ensures that the large difference in flux between continuum and the deep core of telluric absorption lines does not bias the PCA towards either. Since one of the main goals of the PCA is to capture the contribution from the telluric absorption lines and detrend them, we need the spectral channels with telluric lines to be optimally weighted. One option to do so in the context of PCA is to alter the standardization to overweight the telluric lines. However, we choose a comparatively conservative approach which instead weighs all the spectral channels equally by dividing each one of them by their standard deviation. Note that our approach of standardization is not a necessary but an optional step in our PCA detrending approach, and just one of the many ways in which the data can be weighed before performing SVD.

We also mask the spectral channels heavily saturated by telluric absorption and those known to have bad pixels to prevent them from biasing the PCA. We use the same mask for bad pixels applied by \cite{brogi_signature_2012}. For all dates, we mask the first 17 and last 17 pixels at the edges of each detector. In addition, for the second detector, we also mask pixel indices 817 to 830 and 885 to 898 which correspond to a damaged portion of the CRIRES detector.

To perform PCA, we use the singular value decomposition ({\tt svd}) implementation in {\tt numpy} on the standardized data cube to obtain a set of eigenvectors and their corresponding singular values. We use the singular values to rank the eigenvectors in order of the variance they represent, and choose the first \Npca{} eigenvectors for the next steps, where \Npca{} is the number of PCA components used. In Section \ref{sec:compute_optimal_PCA} we describe our process of determining the optimal \Npca{}.  

We then perform the second step of multi-linear regression on the original data cube (before standardization) to compute the best fit linear coefficients (i.e. eigenvalues) for the linear combination of eigenvectors that represents a `fit' to the telluric, instrumental, and stellar variation in the data cube. In our approach, the original eigenvectors were extracted on the standardized data which involves the step of mean subtracting each spectral channel. However, since we intend to detrend the original data cube using these eigenvectors, there is an additional `offset' arising from mean-subtraction step in our standardization that needs to be account for. To do so, we append a `bias vector' of ones (of same the length as an eigenvector) to the eigenvector matrix. We perform multi-linear regression on the original data cube using this matrix of chosen \Npca{} eigenvectors and a bias vector of ones (similar to the approach described in the Methods section of \cite{giacobbe_five_2021}, and in \cite{gibson_relative_2022} in the context of \texttt{SYSREM}. 

We next divide the original data cube by this multi-linear regression model fit. We expect the residual data cube left to be free of the majority of systematics and retaining mainly the planetary signal and noise. An example of the resulting detrended data cube from this step for one of the detector for is shown in the middle panel of Figure \ref{fig:pca_detrending}.

We emphasize that the two additional steps are due to the standardization we choose to implement. An alternative approach of running the SVD on the original datacube would not require the standardization and multi-linear regression as the `fit' or the reconstruction of the the systematics in the data cube can be done using the chosen \Npca{} eigenvectors and their respective eigenvalues.

\subsubsection{Masking insufficiently detrended spectral channels after PCA}
\label{sec:post_PCA_mask}
\begin{figure}
\centering
\includegraphics[width=0.5\textwidth]{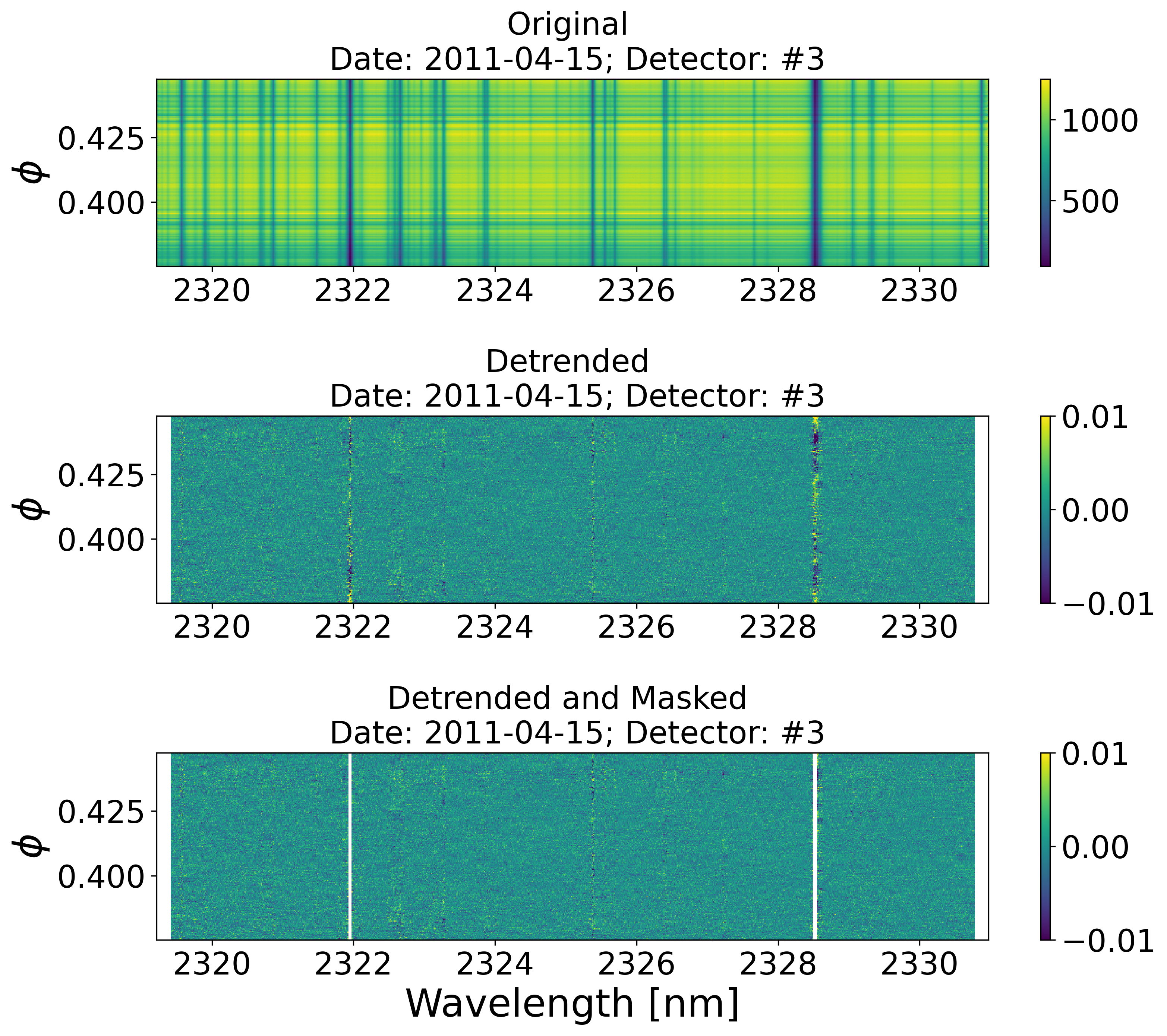}
\caption{Example of PCA detrending for observations taken with detector $\#$3 on the night of 14-04-2011 to 15-04-2011. The top panel shows the original raw flux data cube where each row is the 1D spectrum for an exposure, the horizontal axis is the wavelength, and the vertical axis is the orbital phase for each exposure. The middle panel is the PCA detrended data cube (following the steps described in Section \ref{sec:pca_detrending}) along with bad spectral channels masked out in white. The bottom panel shows the same detrended data cube in the middle panel, with the spectral channels still retaining a significant level of residual noise masked out in white along with the bad pixels. 
}
\label{fig:pca_detrending}
\end{figure}

After PCA, we run an automated routine to check and mask spectral channels that have not been corrected properly in the detrended data cube. This is done on a per date and per detector basis, by checking if the standard deviation for any spectral channel exceeds a threshold value. The threshold value for M spectral channels is computed as N times the median of the standard deviations of all spectral channels. Here, N is the multiple of standard deviations corresponding to the tail probability 1/M for a normal distribution. We compute N using the {\tt scipy.stats.norm.isf()} routine with 1/M as input. This threshold essentially marks the level which, if exceeded by a spectral channel in terms of its standard deviation, is likely due to both high photon noise and insufficient detrending. Such a spectral channel is hence masked and not included in further steps of cross-correlation analysis. The output from this step is shown in the third panel in Figure \ref{fig:pca_detrending}.     

\subsubsection{Optimizing the PCA detrending}
\label{sec:compute_optimal_PCA}
While the PCA detrending is a powerful approach in identifying the dominant contribution from telluric and stellar lines, it requires optimising the \Npca{} chosen. Deciding the optimal \Npca{} is important, as using too few components could result in a suboptimal detrending of the telluric and stellar contamination, while using too many components could result in partial removal of the planetary signal. It is also essential to optimize \Npca{} because the telluric spectrum has absorption lines from \ch{H2O} and \ch{CH4} which are among the main molecular species we are searching for in the planetary spectrum. This implies suboptimal removal of telluric \ch{H2O} and \ch{CH4} lines in the data can lead to retention of telluric contribution to the cross-correlation signal, making it hard to distinguish it from the planetary \ch{H2O} and \ch{CH4}.        
  
Across the literature, there are examples of multiple different approaches taken to optimize the detrending process. Studies using the {\tt SYSREM} method (similar to PCA but with weighting individual pixels by their uncertainties and finding components through iterative fitting) determine the optimal number of iterations as one that maximizes the recovery of the injected planetary signal in the data in form of the cross-correlation function (CCF, see Section \ref{sec:cross_correlation_logL} for more on its computation) value at the known planetary velocity. \cite{cabot_robustness_2019} suggest that this approach could lead to overfitting, and instead recommend optimizing the injected signal at a range of planetary velocities. More recent works \citep{spring_black_2022, holmberg_first_2022} have determined the optimal \Npca{} as one that minimizes the difference between observed CCF and injected CCF (referred to as $\Delta$CCF), making the optimization independent to the injected model itself. \cite{cheverall_robustness_2023} performed a comparison of these methods and find that optimizing based on $\Delta$CCF is less likely to be biased, i.e. affected by systematic amplification in S/N of the retrieved signal, as compared to when optimizing based on injected CCF. 

All the aforementioned methods take the approach of maximizing the planetary signal to determine the optimal \Npca{}. An alternative, more conservative approach is to minimize contributions to the CCF from telluric \ch{H2O} and \ch{CH4} absorption lines in a range of velocities around 0 km/s (i.e., the observer's rest frame) where the strongest contribution from the telluric absorption lines (stationary with respect to the observatory) is expected \citep{lafarga_hot_2023}. This approach might not result in the strongest planetary signal, but it accomplishes the main aim of PCA, which is suppressing telluric lines. In this work, we follow the same approach as \cite{lafarga_hot_2023} and determine the optimal \Npca{} as the minimum \Npca{} needed to minimize the contribution to the CCF from telluric lines. As an extension to this method, we define a new metric that can be used to measure the telluric contribution to the CCF for each \Npca{}. Note that in comparison to this approach, the methods based on maximizing the CCF or $\Delta$CCF used by previous works e.g. \cite{cheverall_robustness_2023} are statistically biased because using that method one always optimizes the local noise together with the signal, and not only the signal. This is statistically incorrect and may result in noise amplification for each of the orders optimized, thus always leading to artificially higher detections. Optimizing the $\Delta$CCF is unbiased in terms of recovering the optimal planet signal, but the noise is cancelled out, so it does not allow us to estimate whether the telluric contribution is appropriately suppressed. 

\begin{figure*}
\centering
\includegraphics[width=0.8\textwidth]{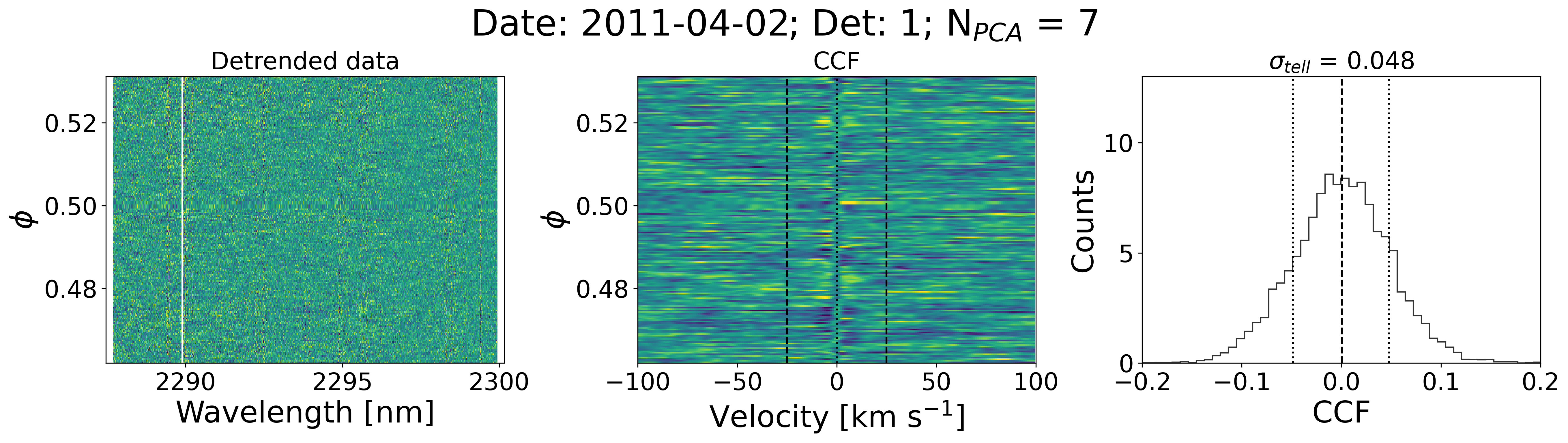}
\caption{Demonstration of the steps we follow to determine the choice of \Npca{} that leads to optimal suppression of telluric contribution to the CCF for one of the detectors from one night of observation. In the left panel, we show the PCA detrended data cube (Y axis showing the phase $\phi$ for each exposure) along with the noisy spectral channels masked in white. In the middle panel, we show the CCF of the data with an ESO SkyCalc model for each phase for a range of velocities, with the dashed lines showing the $\pm$25 \kms{} region around the dotted line marked by 0 \kms{}. In the third panel, we show the distribution of the CCF values in the $\pm$25 \kms{} region around 0 \kms{}, with the dashed lines marking the $\pm$1$\sigma$ of the distribution (referred to as \sigmatell{}) around the mean of the distribution at 0. }
\label{fig:N_PCA_optimization_demo}
\end{figure*}

We compute forward models for the telluric transmission flux using ESO SkyCalc at resolution R = 100000 for CRIRES. This model accounts for absorption from the dominant molecules in the telluric spectrum, including \ch{H2O} and \ch{CH4}. For a range of values of \Npca{} from 2 to 15 in increments of 1, we detrend the data cube and mask insufficiently the detrended spectral channels for each detector and for each date as described in Section \ref{sec:pca_detrending}. We then compute the CCF between the observed spectrum from each exposure with the \ch{H2O} model Doppler shifted by velocities ranging from -100 to 100 \kms{} in steps of 1 \kms{}. This yields a CCF trail matrix with dimensions (time $\times$ velocity) as shown in Figure \ref{fig:N_PCA_optimization_demo} and Figure \ref{fig:N_PCA_optim_trail_matrix_histogram}for one of the detectors for one date of observation. 

Telluric \ch{H2O} in the data is expected to vary during the course of exposures taken during one night, which can lead to a correlation or anti-correlation with the telluric ESO SkyCalc model computed for a single value of precipitable water vapour. This respectively appears as a positive or negative values in the CCF trail matrix. Since we are interested in evaluating the absolute contribution from telluric water in the CCF trail matrix, we compute the average of the absolute CCF values along the time axis. We then mean subtract this average CCF and divide by the standard deviation across the whole velocity range, shown in Figure \ref{fig:N_PCA_optim_trail_matrix_histogram}. Since the telluric absorption lines are nearly stationary in the observatory's rest frame, we expect their contribution to the CCF to be highest around 0 \kms{}. For a given \Npca{}, significant peaks around 0 \kms{} in this 1D CCF indicate the presence of uncorrected telluric contribution in the data, implying that the particular \Npca{} is suboptimal. However, it is also possible that the telluric contribution from the CCF is not localized in one peak exactly at 0 km/s, and can instead be spread across a range of velocities near 0 km/s. Simply averaging the CCF values along the time axis and considering the peak value of the average CCF at 0 \kms{} loses information on the non-zero telluric contribution to the CCF spread in time and velocity. We need an alternative metric that encodes the statistical variation of CCF in the ranges of velocities around 0 \kms{} for all exposures.       

\begin{figure*}
\centering
\includegraphics[width=0.8\textwidth]{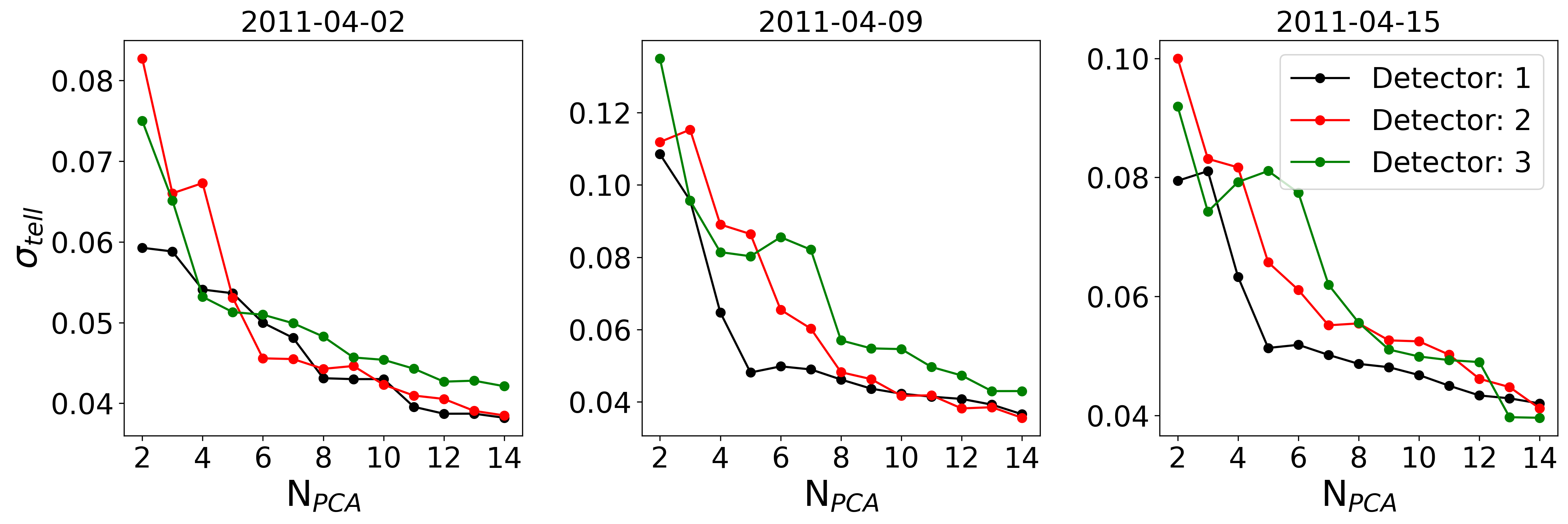}
\includegraphics[width=0.8\textwidth]{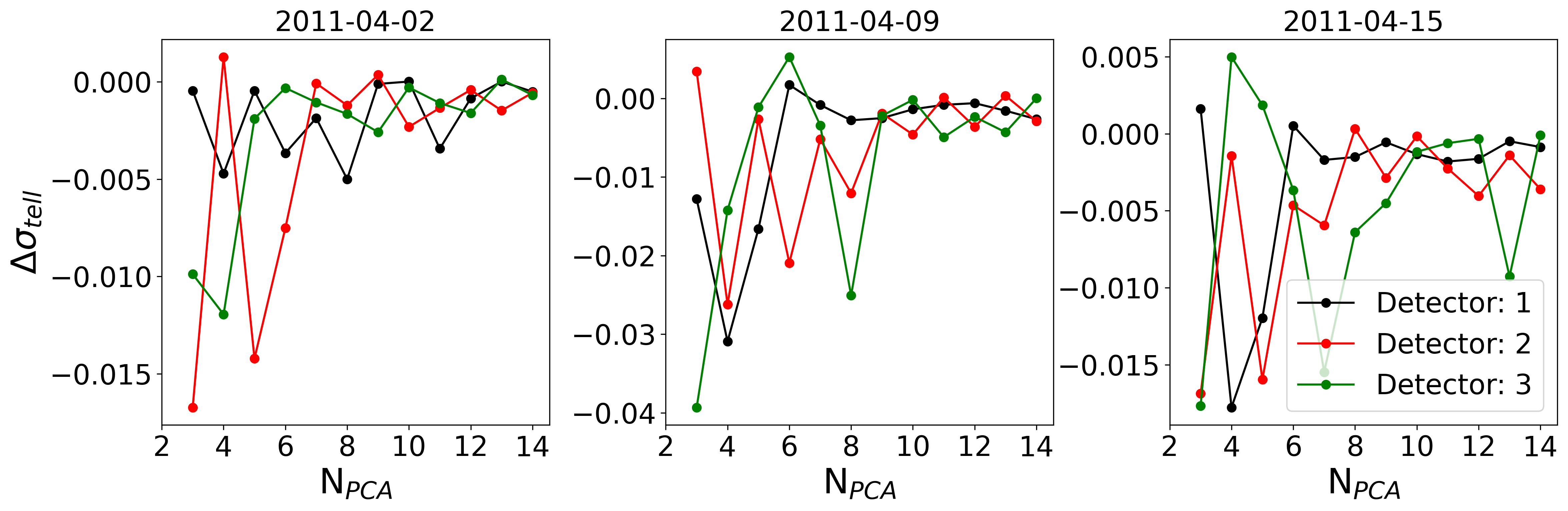}
\caption{Variation of \sigmatell{} (top three panels) and $\Delta$\sigmatell{} (bottom three panels) with \Npca{} for all detectors and dates as described in Section \ref{sec:compute_optimal_PCA}. We expect that the optimal suppression of telluric contamination in the data with  changing \Npca{} will diminish the amplitude of the CCF of the data with the ESO SkyCalc model. This will cause the distribution of CCF values around 0 \kms{} to become narrower, which makes \sigmatell{} decrease with \Npca{}. Broadly for all dates and detectors, the incremental improvement with \Npca{} indicated by $\Delta$\sigmatell{} plateaus around \Npca{} = 9, which we choose as the optimal \Npca{}.}
\label{fig:N_PCA_optim}
\end{figure*}

To devise such a metric, we define a region of $\pm$25 \kms{} centred on 0 \kms{} and compute the histogram of the CCF values for all the exposures in this velocity range, shown in Figure \ref{fig:N_PCA_optimization_demo} for a given value of \Npca{}. We take the average of $+1\sigma$ and $-1\sigma$ value of this histogram ($+34\%$ and $-34\%$ quantile from the median) as our metric (from now on referred to as \sigmatell{}) to decide how well the PCA detrending of the telluric contribution has performed with respect to changing \Npca{}. A smaller \sigmatell{} value implies a narrower distribution of CCF which means relatively more CCF values are close to zero, meaning better removal of telluric contribution and vice versa. We calculate \sigmatell{} on a per detector and per data basis for \Npca{} values ranging from 2 to 14 in steps of 1. To assess the improvement with change in \Npca{}, we also calculate the incremental change in \sigmatell{} as $\Delta $\sigmatell{}. We show the variation of \sigmatell{} and $\Delta $\sigmatell{} with \Npca{} in Figure \ref{fig:N_PCA_optim}. 

Across all three detectors for all three dates, we find \sigmatell{} broadly decreasing with increasing \Npca{} as expected. very night and every detector behaves differently, for instance Night 2 detector 1 flattens after 5 components, whereas on the same night detector 2 flattens around 8. For all cases, the \sigmatell{} continues to decrease even after 8–9 components. Therefore, simply from the variation of \sigmatell{} it is not possible to objectively decide on the optimal \Npca{}. However, when considering the trend in $\Delta $\sigmatell{}, which shows the incremental change in \sigmatell{} with \Npca{}, we find that $\Delta $\sigmatell{} plateaus around \Npca{} = 9 for all detectors for all dates. Hence, we choose \Npca{} = 9 as the optimal value for all further analysis in this work. 

We caution the reader that our metric in the form of \sigmatell{} joins the list of the metrics used in the literature so far to decide the optimal intensity of telluric detrending procedure. We refrain from claiming this is the best metric for optimizing the PCA or {\tt SYSREM} as that would require a detailed investigation and comparison of all the metrics using injection and recovery tests which is beyond the scope of this work.

\subsection{Cross-correlation spectroscopy}
\label{sec:cross_correlation_spectroscopy}
After performing the optimal detrending of the data, we move to searching for the signal from the planetary atmosphere using cross-correlation. In this section, we describe the formalism we use to compute the cross-correlation between a model of the planetary atmosphere and the data. We also describe the procedure we follow to perform the model reprocessing necessary to account for the detrimental effects of the PCA detrending on the planetary signal present in the data.   

\subsubsection{Model calculation}
\label{sec:model_calculation}
We compute the emission spectrum model of \taubootisb{} using {\tt GENESIS} \citep{gandhi_genesis_2017, pinhas_h2o_2019}. {\tt GENESIS} performs a line-by-line numerical radiative transfer calculation, yielding a spectrum of the planet for a given atmospheric temperature and chemical abundance profile. We then divide this spectrum by a blackbody spectrum of the star to obtain the observed emission spectrum of the planet as the ratio of planetary flux to stellar flux (\FpFs{}). {\tt GENESIS} computes the opacity of each species on a grid of pressure-temperature ($P$-$T$) values corresponding to a given $P$-$T$ profile. The $P$-$T$ profile is parametrized by the values of two ($P$-$T$) pairs : ($P_1, T_1$) and ($P_2, T_2$) where $P_1>P_2$. We use a range of fixed pressure values, between 100 bars to 10$^{-8}$ bars. For pressures lower than $P_1$ and higher than $P_2$ we take the $P$-$T$ profile to be isothermal. For pressures between $P_1$ and $P_2$, the $P$-$T$ profile is linear with respect to log P. Our priors on $P_1$ and $P_2$ when performing the retrieval allow for both inverted and non-inverted profile, with the only condition imposed being $P_1>P_2$.

The chemical abundances of each species under consideration, which includes \ch{H2O}, \ch{CH4}, \ch{CO}, \ch{NH3}, \ch{HCN}, and \ch{CO2}, are considered to be vertically well-mixed. The assumption of vertically well-mixed abundances is predicted by both equilibrium and non-equilibrium chemistry models \cite{moses_chemical_2013} for CO and water \ch{CO} and \ch{H2O} which are the two main molecules of interest in the wavelength range covered by the data. The opacity data we use are based on the latest line-lists available for each molecule: \cite{polyansky_exomol_2018} for \ch{H2O}, \cite{rothman_hitemp_2010, li_rovibrational_2015} for \ch{CO}, \cite{hargreaves_accurate_2020} for \ch{CH4}, \cite{coles_exomol_2019} for \ch{NH3}, \cite{harris_improved_2006, barber_exomol_2014} for HCN, and \cite{huang_semi-empirical_2013, huang__ames-2016_2017} for \ch{CO2}. We compute the model over a wavelength grid corresponding to a constant resolving power of 250000 within 2.25 to 2.35 micron. Before computing the cross-correlation, we convolve the model to the resolution of CRIRES using a Gaussian kernel with FWHM corresponding to the resolution of 100000 for CRIRES. The convolution is done with a Gaussian kernel prior to regridding the model to the wavelength solution of the data. We neglect the changes in instrumental resolution across the narrow wavelength range of the CRIRES data. Changes to the resolution across the detectors are possible but not documented in literature or measurable from the data themselves.

\subsubsection{Cross-correlation and mapping to log-likelihood space}
\label{sec:cross_correlation_logL}

To evaluate the statistical significance of a detection, we choose to use the cross-correlation to log-likelihood framework \citep[C-to-$\log L$,][]{brogi_retrieving_2019}. This framework allows us to map the cross-correlation between the data and the model spectrum to a likelihood probability, which can then be used for confidence interval estimation and Bayesian inference. Estimating the Bayesian evidence through this approach can also be used for model comparison. 

This cross-correlation to log-likelihood mapping, as prescribed in \citep{brogi_retrieving_2019}, is:

\begin{equation}\label{eq:cc2logL}
\log(L) = - \frac{N}{2} \log [ s^2_f - 2R + s^2_g ],
\end{equation}

where $s^2_f$ is the variance of the observed spectrum, $s^2_g$ is the variance of the emission spectrum model, $R$ is the cross-covariance between the observed spectrum and model (Doppler shifted by a given RV), and $N$ is the total number of points in the spectrum for an exposure. We compute the log-likelihood separately for each detector. For each detector, we compute the log-likelihood for each the part of the spectrum on that detector for each time instance, and sum them across time to obtain the total log-likelihood for the detector. We then sum the log-likelihood across all the detectors to obtain the total log-likelihood for a given date. Note that is important to clarify how the log-likelihood calculation is partitioned across the detectors as this can have implications depending on how the PCA detrending might be done differently across detectors.

The cross-correlation is naturally part of this formalism, since the correlation coefficient $C$ is proportional to the cross-covariance $R$ as: 

\begin{equation}\label{eq:r2cc}
{C} = \frac{R}{\sqrt{s^2_f s^2_g}}.
\end{equation}

We compute the cross-covariance $R$ between a model spectrum Doppler shifted by a given RV and the detrended observed spectrum per exposure for each detector and each date. We then normalize it by the product of the model and data variance, which yields one scalar value of $C$ per exposure, detector, and date for a given RV. Going forward, we refer to $C$ as CCF. To be clear, the equations here are for illustrating the relation between $C$ and $R$. Practically, we compute $R$, $s^2_f$, and $s^2_g$ first, and combine them to compute the log-likelihood as per Equation \ref{eq:cc2logL}, and the cross-correlation coefficient $C$ as per Equation \ref{eq:r2cc}.  

The total RV for an exposure is computed using the corresponding planetary orbital phase. Assuming a circular orbit, it takes the following form: 

\begin{equation}
\label{eq:rvcalc}
\mathrm{RV}(\phi) = K_\mathrm{P} \mathrm{sin} (2\pi \phi) + V_\mathrm{sys} + V_\mathrm{bary}(\phi) 
\end{equation}   

where RV is the radial velocity shift from the planet, \Kp{} is the Keplerian velocity amplitude of the planet, $\phi$ is the orbital phase, \Vsys{} is the systemic radial velocity of the system with respect to the barycentre of the solar system, and $V_\mathrm{bary}(\phi)$ is the barycentric earth radial velocity (BERV) computed for the observing site Paranal for each time step of observation (hence a function of $\phi$). 

A single pair of (\Kp{}, \Vsys{}) value gives one set of RVs for all the exposures through the Equation \ref{eq:rvcalc}. For a given date and detector, we compute the CCF for each exposure by Doppler shifting the model spectrum according to the RV for that exposure. Summing across exposures, and then across the detectors and dates, we obtain the total CCF for a single pair of (\Kp{}, \Vsys{}). Analogously, the total log-likelihood for a model can be calculated using the $C$-to-$\log L$ map in Equation \ref{eq:cc2logL}. Doing this for a grid of (\Kp{}, \Vsys{}) values yields the familiar 2D \KpVsys{} maps for CCF and log--likelihood seen in typical HRCCS works. However, an accurate estimation of the log-likelihood requires additional reprocessing of the planetary model, as we describe in the next section.    

\subsubsection{Model reprocessing}
\label{sec:model_reprocessing}
For an accurate computation of the log-likelihood values, it is also important to take into account the effect the PCA detrending has on the observed planetary signal. For the orbital phases covered by our observations, we expect the planetary lines to be Doppler shifted by $\sim$230 \kms{} from beginning to end of the observations for each night. This corresponds to a large shift of $\sim$150 pixels on the detector. Despite the PCA approach being primarily sensitive to modeling the strong telluric and stellar lines that are stationary as compared to the planetary lines, it can also affect the signal from the planetary atmosphere in the data. Hence, it is necessary to reproduce the effect of PCA on the model spectrum used for cross-correlation. To do so, we follow the steps of model reprocessing, which have previously been also followed by \cite{brogi_roasting_2023}.

\begin{figure}
\centering
\includegraphics[width=0.5\textwidth]{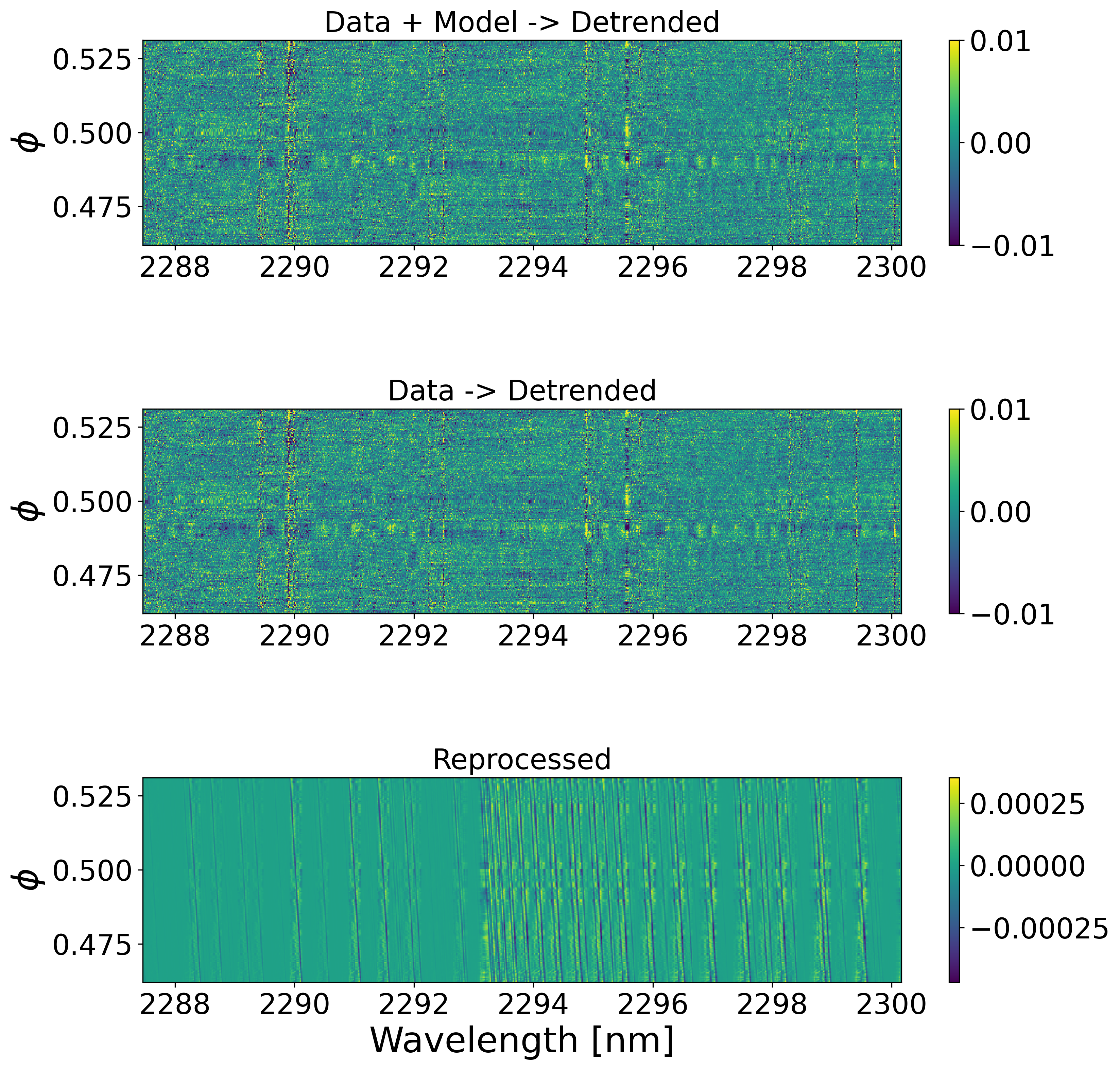}
\caption{Model reprocessing for detector 1 on date 02-04-2011 following the steps described in Section \ref{fig:model_reprocessing}. The top panel shows the original data cube with the {\tt GENESIS} model injected and detrended using steps in Section \ref{sec:pca_detrending} using the same eigenvectors computed using PCA on the data cube itself (middle panel). The bottom panel shows the reprocessed model cube obtained after subtracting the data cube in the middle panel from the top panel. The effect of PCA on the planetary signal in the data is reproduced in the reprocessed model cube and allows for a more accurate computation of cross-correlation. 
}
\label{fig:model_reprocessing}
\end{figure}

The steps of model reprocessing are illustrated in Figure \ref{fig:model_reprocessing}. We first run the PCA on a data cube for each detector (following the same steps for all dates) and obtain the detrended data cube and the PCA eigenvectors. For a given \KpVsys{} pair of values, we obtain the total RV for each phase using Equation \ref{eq:rvcalc}. We apply Doppler shift to a given model spectrum by the RV for each phase. We then inject it in the observed data cube to obtain a model injected data cube with the same dimensions as the observed data cube. 

The same eigenvectors used for detrending the respective observed data cube are then used to perform multi-linear regression on the model injected data cube, and the fit is divided out from the model injected data cube. Hence, the residual obtained is the detrended model injected data cube. We next subtract the detrended data cube from the detrended model injected data cube, to obtain the reprocessed model cube. 

This reprocessed model cube is eventually used to compute the cross correlation and log likelihood per exposure, which is then summed across all exposures to obtain the total cross correlation and log likelihood for the respective detector for the respective date (and for the given pair of \KpVsys{} values) as described in Section \ref{sec:cross_correlation_logL}.

\subsection{Atmospheric retrieval setup} 
\label{sec:atmospheric_retrieval_setup}
With a framework in place that calculates accurate log-likelihood values for a given forward model, the next step is to couple this with a Bayesian retrieval framework. This involves using a probability distribution sampling algorithm, e.g. using an MCMC, to sample the log-likelihood distribution for a range of parameters that an atmospheric model can be parametrized by. In this work, we parametrize a model {\tt GENESIS} emission spectrum by the volume mixing ratios (VMR) of various species, the $P$-$T$ profile, and the (\Kp{},\Vsys{}) values that determine the total RV by which the model is Doppler shifted. We use the {\tt MultiNest} algorithm \citep{feroz_multinest_2009} implemented in {\tt pymultinest} by \cite{buchner_x-ray_2014} for sampling the log-likelihood distribution and computing the marginalized posterior distributions for model parameters. Nested sampling using {\tt pymultinest} is particularly advantageous as compared to other MCMC based sampling methods as the former can also calculate model evidence which can be used for model comparison.

We show the fixed model parameters and the free model parameters along with their priors in Table \ref{tab:param_priors_best_fit}. Despite being a non-transiting planet, the orbital solution for \taubootisb{} is well-known from previous works. We chose our priors to be wide enough to allow for potential atmospheric effects that could cause a shift in the velocity at which the planetary signal is detected. For molecular abundances, lower bound corresponds to insignificant contribution to the signal, and the upper bound ensures that the planet is still in a hydrogen dominated atmosphere regime (which is expected for a planet of this mass). The $P$-$T$ profile parameters are ad-hoc to be wide enough and allow for both inverted and non-inverted $P$-$T$ profiles. We include a multiplicative scaling factor ($f_s$) for the model to account for the uncertainty on the radius of \taubootisb{}. The prior on log$f_s$ is ad hoc and wide enough to allow for appropriate scaling due to unknown planetary radius and stellar photosphere level. 

We run the retrieval with the number of live points set to 2000, and a tolerance $\Delta$lnZ (Z is the Bayesian evidence) set to 0.5. The retrieval converges when the tolerance on $\Delta$lnZ is reached, resulting in $\sim$20000 evaluations of the log-likelihood. We checked and ensured that changing the number of live points did not affect the convergence of the retrieval, obtaining similar final results in terms of the parameter estimates and Bayesian evidence when using different number of live points.

\begin{table*}
    \centering
    \begin{tabular}{cccc}
        \hline
        \hline
        &Fixed Parameters&\\
        \hline
         Parameters & Fixed value & Reference\\
         \hline
         R$_{\mathrm{P}}$ [\Mjup{}]&  1.25& estimated$^{*}$\\
         log$_{10}g_{\mathrm{P}}$ [cgs]& 4.013 & estimated$^{*}$\\
         R$_{*}$ [\Msun{}]& 1.42 $\pm$ 0.08 & \cite{borsa_gaps_2015}\\
         T$_{*}$ [K]& 6399 $\pm$ 45 & \cite{borsa_gaps_2015}\\
         $e$& 0 & within 1$\sigma$ of \citep{rosenthal_california_2021}\\
         T$_{0}$ [HJD] & 2455652.108 $\pm$ 0.004 & \cite{brogi_signature_2012}\\
        \hline
        \hline
        \\
        \hline
        \hline
        &Free Parameters&\\
        \hline
         Parameters & Prior & Median and $\pm$1$\sigma$ & MAP\\
         \hline
         \logfs{}& $\mathcal{U}$[-4, 1] & --0.61$^{+0.36}_{-0.3}$&--0.65\\
         \Kp{} [\kms{}]& $\mathcal{U}$[100, 120] & 110.92$^{+0.51}_{-0.49}$&110.73\\
         \Vsys{} [\kms{}]& $\mathcal{U}$[-20, 0,] & --17.18$^{+0.26}_{-0.26}$&--17.11\\
         log$_{10}$P$_1$ [Pa]& $\mathcal{U}$[-3, 7] & 5.31$^{+1.11}_{-1.8}$&5.63\\
         T$_1$ [K]& $\mathcal{U}$[300, 3000] & 2312.31$^{+394.57}_{-456.98}$&2447.46\\
         log$_{10}$P$_2$ [Pa]& $\mathcal{U}$[-3, 7]  & --0.29$^{+2.09}_{-1.75}$&3.57\\
         T$_2$ [K]& $\mathcal{U}$[300, 3000]  & 1270.65$^{+544.23}_{-535.41}$&1880.27\\
         \logCO{} [VMR]& $\mathcal{U}$[-15, -0.5] & --3.43$^{+1.62}_{-0.86}$&--4.2\\
         \logWater{} [VMR]& $\mathcal{U}$[-15, -0.5] & --5.14$^{+1.23}_{-6.38}$&--4.67\\
         \logMethane{} [VMR]& $\mathcal{U}$[-15, -0.5] & --8.65$^{+4.0}_{-4.07}$&--5.02\\
         \logAmmonia{} [VMR]& $\mathcal{U}$[-15, -0.5]  & --9.7$^{+3.4}_{-3.42}$&--5.79\\
         \logCarbondiox{} [VMR]& $\mathcal{U}$[-15, -0.5] & --8.51$^{+4.27}_{-4.11}$&--5.1\\
         \logHydrogencyn{} [VMR]& $\mathcal{U}$[-15, -0.5] & --9.73$^{+3.4}_{-3.35}$&--6.37\\
        \hline
        \hline
        \\
        \hline
        \hline
        &Derived Parameters&\\
        \hline
        $i[^{\circ}]$ & & 43.92$\pm$0.52 \\
        M$_{\mathrm{P}}$[\Mjup{}] & & 5.95 $\pm$ 0.28 \\
        \hline
        \hline
        &Atmospheric composition&\\
        \hline
            & & Free chemistry & Equilibrium chemistry\\
        \hline
        C/O & & 0.95$^{+0.06}_{-0.31}$ & 0.75$^{+0.26}_{-0.21}$\\
        M/H [dex]&&--0.21 $^{+1.66}_{-0.87}$ & --0.69$^{+1.29}_{-0.62}$\\
        \hline
        \hline
    \end{tabular}
    \caption{Fixed (first table) and free retrieval parameters and their priors (second table), derived parameters (third table), and inferred atmospheric composition (fourth table) for \taubootisb{} and the host star in this work. $\mathcal{U}$ denotes a uniform prior within the range specified. The column `MAP' refers to the maximum a posteriori solution obtained from the retrieval.} $^{*}$The radius and gravity of \taubootisb{} is uncertain because the planet is non-transiting, and is estimated based on the known hot-Jupiters similar in mass as \taubootisb{}.
    \label{tab:param_priors_best_fit}
\end{table*}
\section{Results}
\label{sec:results}
We describe the main results from our analysis in this section, followed by a discussion of their implications in Section \ref{sec:discussion}. 
\subsection{Best fit model parameters from retrievals}
\label{sec:results_best_fit_model_params}
We list the best fit model parameters from the retrieval with \Npca{} fixed to 9 in Table \ref{tab:param_priors_best_fit}. The full corner plot for the posterior distributions for all the free parameters are shown in Figure \ref{fig:retrieval_posteriors_full}. In the next subsections, we focus on the results for selected parameters that have the tightest constraints and are useful for our interpretation of \taubootisb{}'s atmosphere.

\subsubsection{Constraints on \Kp{} and \Vsys{}}
\label{sec:results_KpVsys_constraints}
We measure \Kp{} = 110.9 $\pm$ 0.5 \kms{} from the retrieval, which is consistent within 1$\sigma$ with the previous measurement \Kp{} = 110.2 $\pm$ 3.2 \kms{} from this dataset by \cite{brogi_signature_2012}. Our measurement is also consistent within 1$\sigma$ with measurements from other instruments including Keck-NIRSPEC \citep[111 $\pm$ 5 \kms{}, ][]{lockwood_near-ir_2014} and SPIRou \citep[109.2 $\pm$ 0.4 \kms{}, ][]{pelletier_where_2021}. However, our constraint on \Kp{} is offset by $\sim$3$\sigma$ from the value 106.21$^{+1.76}_{-1.71}$ \kms{} retrieved by \cite{webb_water_2022} from CARMENES observations. 

We measure \Vsys{} = -17.2 $\pm$ 0.2 \kms{}, which is offset by $\sim$3$\sigma$ from the value inferred from the same dataset by \cite{brogi_signature_2012} (-16.4 $\pm$ 0.1 \kms{}). However, it is important to note that the \Vsys{} value quoted by \cite{brogi_signature_2012} is not measured, but rather imposed by extrapolating from the previously measured value by \cite{donati_magnetic_2008} and adjusting the time of inferior conjunction $T_{0}$ accordingly. We emphasize that we are using the same wavelength solution for the here as from the first work on this dataset by \cite{brogi_signature_2012} which is found to be precise at the level of 100 m/s (an order of magnitude smaller than the offset to \Vsys{} that we measure). The time of inferior conjunction was refined by \cite{brogi_signature_2012} to match the known radial velocity of the star ( -16.4 \kms{}). Hence, an uncertainty in \Vsys{} may reflect the uncertainty on time of inferior conjunction.

Our \Vsys{} constraint is also offset by $\sim$9$\sigma$ from the value measured by \cite{pelletier_where_2021} from SPIRou (-15.4 $\pm$ 0.2 \kms{}) and by $>$9$\sigma$ from the values measured by \cite{webb_weak_2020} from CARMENES (-11.5 $\pm$ 0.6 \kms{}). A possible reason for these offsets from the SPIRou and CARMENES measurements could be the RV shift induced by the wide-orbit binary companion \taubootisStar{} B \citep{justesen_constraining_2019}.  

\begin{figure}
\centering
\includegraphics[width=0.5\textwidth]{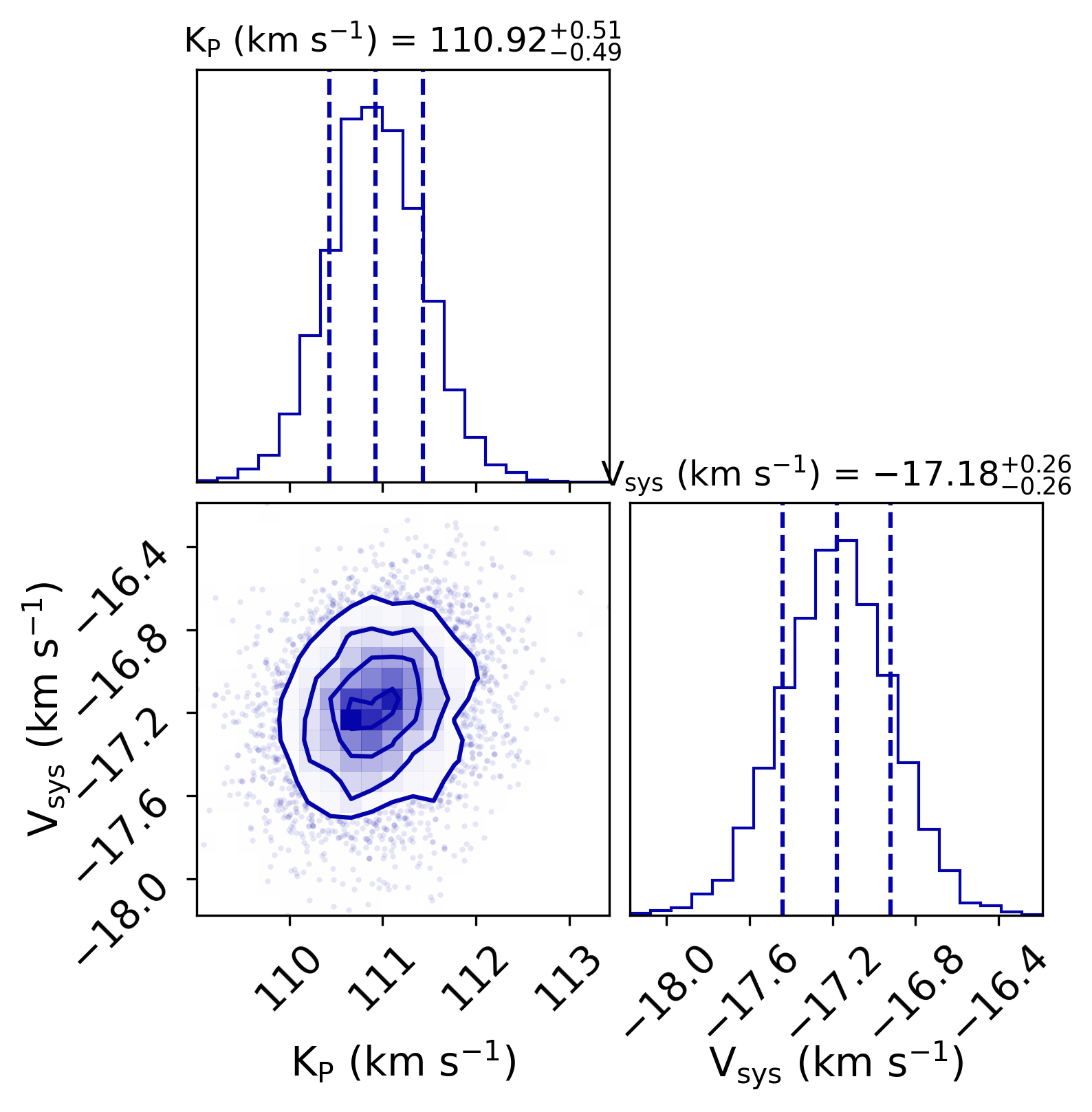}
\caption{Posterior distributions of the radial velocity semi-amplitude, \Kp{} and the systemic offset velocity, \Vsys{} from the joint retrieval analysis of all three nights of data.}
\label{fig:KpVsys_posteriors}
\end{figure}

We obtain remarkable precision of $0.5$ \kms{} and $0.3$ \kms{} precision on the \Kp{} and \Vsys{} respectively as shown in Figure \ref{fig:KpVsys_posteriors}. We use these \Kp{} and \Vsys{} measurements to update the constraints on the orbital inclination and mass of \taubootisb{}. Considering the velocity and mass ratios for the planet and the star, we know that K$_{*}$/\Kp{} = M$_{\mathrm{P}}$/M$_{*}$. The radial velocity amplitude for the host star is K$_{*}$ = 0.47$\pm$0.002 \kms{} \citep{justesen_constraining_2019}, stellar mass is M$_{*}$ = 1.35 $\pm$ 0.03 \Msun{} \citep{takeda_structure_2007}. Using these values, we estimate the mass of \taubootisb{} to be M$_{\mathrm{P}}$ = 5.97 $\pm$ 0.14 \Mjup{}. Assuming a circular orbit and using the orbital parameter constraints from \cite{rosenthal_california_2021} on the semi-major axis $a$ = 0.049 $\pm$ 0.0004 AU and orbital period P$_{\mathrm{orb}}$ = 3.312 $\pm$ 5.7 x 10$^{-6}$ days, we estimate inclination of the planet as $i$ = sin$^{-1}$(\Kp{}/(2$\pi a$/P$_{\mathrm{orb}}$)) = 43.92$\pm$0.52 $^{\circ}$. Our mass and inclination estimates are consistent with estimates from the same dataset by \cite{brogi_signature_2012} (M$_{\mathrm{P}}$ = 5.95 $\pm$ 0.28 \Mjup{}), and with the estimates from other datasets by \citep{webb_water_2022} (M$_{\mathrm{P}}$ = 6.24 $^{+0.17}_{-0.18}$ \Mjup{}) and \cite{pelletier_where_2021} (M$_{\mathrm{P}}$ = 6.24 $\pm$ 0.23 \Mjup{}) within 3$\sigma$ uncertainties. 

Assuming the planet is tidally locked, we calculated the projected equatorial rotational velocity for the orbital inclination of 43.92$\pm$0.52 $^{\circ}$ of the planet, which amounts to 1.36$\pm$0.73 \kms{}. If the signal was coming from only the equatorial limb of the planet this would indeed be sufficient to explain the measured shift in \Kp{} and \Vsys{}. However, this configuration would be hard to maintain across the large range of orbital phases covered by the observations which probe the signal originating from across the visible planetary disc.

\subsubsection{Constraints on the atmospheric model}
\label{sec:results_atmospheric_model}
\begin{figure*}
\centering
\includegraphics[width=0.8\textwidth]{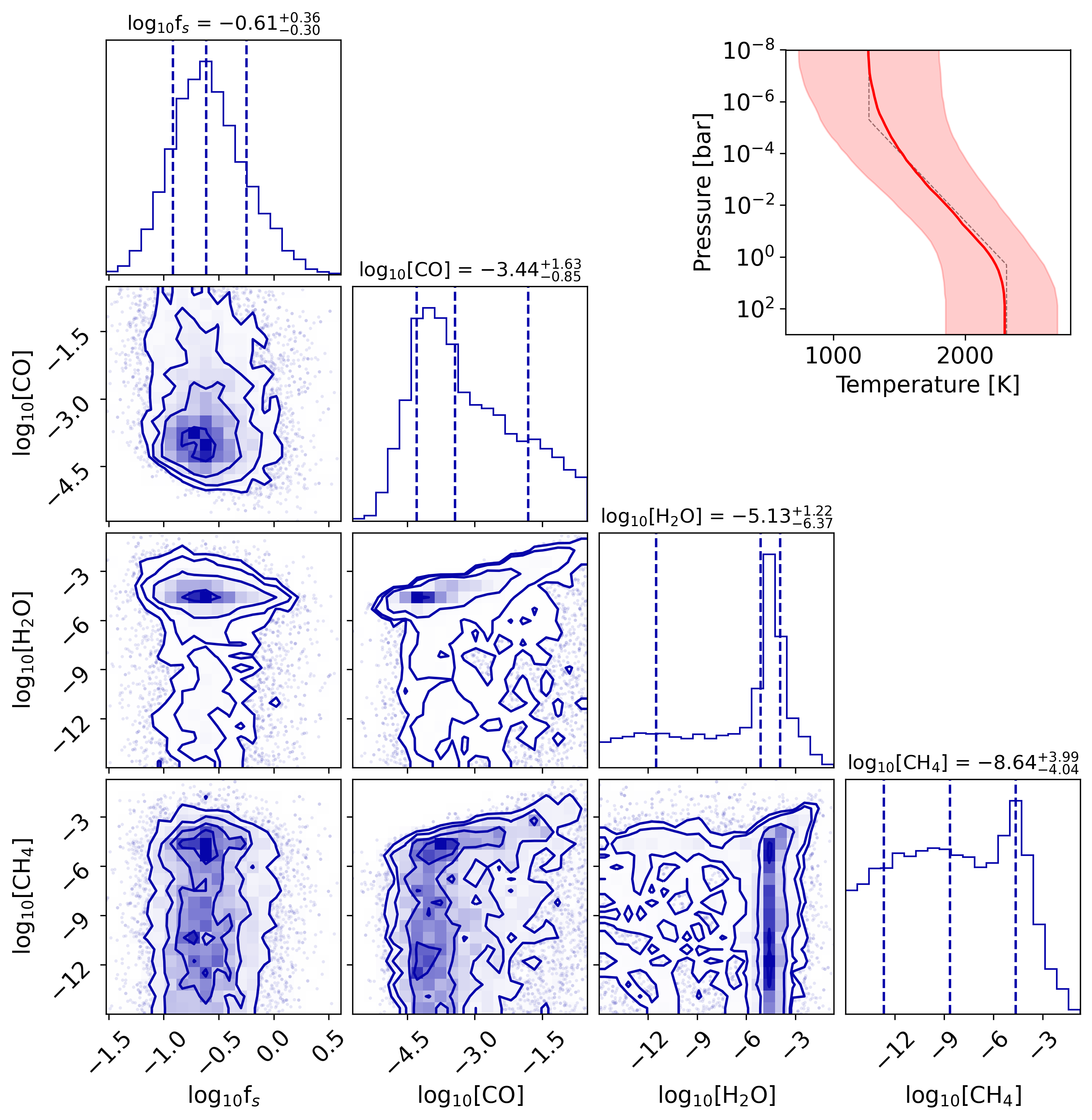}
\caption{Posterior distributions for the model scaling factor and abundances (in log$_{10}$ of volume mixing ratio) of selected molecular species (CO, \ch{CH4}, and \ch{H2O}) included in the joint retrieval of all three nights of the data. The top right panel shows the median (solid red) and $\pm1\sigma$ (shaded red) constraint on the temperature pressure (TP) profile of the atmosphere. The black dashed \TP{} profile is computed corresponding to the median parameters of the \TP{} profile obtained from the retrieval.}
\label{fig:abundance_posteriors}
\end{figure*}

We show the posterior distributions for selected parameters for the emission spectrum model in Figure \ref{fig:abundance_posteriors}. We find the model scaling parameter to be $f_s$ = 0.24 $\pm$ 0.17 which is indicative of our overestimation of the planetary radius which we fixed to 1.25 \Rjup{}. In addition to this, the scaling factor significantly less than unity could indicate missing sources of opacities (unlikely in this case), or we might have normalized the emission spectrum in units of stellar flux (F$_{\mathrm{S}}$) incorrectly (when computing \FpFs{}). An example of this could be if the stellar effective temperature or radius chosen do not correctly reproduce the stellar spectrum around 2.3 micron. 
Through the retrieval we infer a non-inverted \TP{} profile as shown in Figure \ref{fig:abundance_posteriors} with the tightest constraint in the range of 10$^{-1}$ to 10$^{-3}$ bars. The non-inverted nature of the profile indicates the presence of molecular absorption features, including CO which is the main driver of the detections here, as we show further in Section \ref{sec:KpVsys_maps}. 

We find significant evidence for the presence of CO seen as a clear peak in the posteriors, with VMR \logCO{} = -3.44 $^{+1.63}_{-0.85}$ which is consistent within 1$\sigma$ with the expected solar abundance value \logCO{} $\sim$ -3.3 for equilibrium chemistry \citep{madhusudhan_co_2012, moses_chemical_2013}. We do not find significant evidence for \ch{CH4}, \ch{NH3}, \ch{CO2}, and \ch{HCN} as shown in Figure \ref{fig:retrieval_posteriors_full}. For the abundance of \ch{H2O}, we find a peak in the posterior at VMR \logWater{} = -5.13 accompanied by a non-negligible tail towards lower abundances. The peak as well as the upper 3$\sigma$ limit at \logWater{} = -1.12 found by our retrieval is significantly higher than the upper 3$\sigma$ value of \logWater{} = -5.66 reported by \cite{pelletier_where_2021}. In summary, we are only marginally able to constrain the abundance of \ch{H2O}.

We quantify if the inclusion of \ch{H2O} significantly improves the likelihood by comparing the log-evidence (log$Z$) obtained from {\tt pymultinest} for the retrieval that includes \ch{H2O}, and another retrieval with identical setup except for the inclusion of \ch{H2O} in the model. Based on the log Z estimated from these two retrievals, we find that the model with \ch{H2O} is preferred very marginally at 1.5$\sigma$ confidence (corresponding to $\Delta$log$Z$ = 0.3) over a model without \ch{H2O}. 

However, this approach of model selection has shortcomings because the Bayesian evidence can be sensitive to the priors used for the free parameters. In the context of this work, we expect the log$Z$ to be dependent on the prior range of \ch{H2O}. We perform a simple test for sensitivity of log$Z$ on the specified prior for \ch{H2O} abundance by running two additional retrievals: one with narrower prior on \logWater{}: $\mathcal{U}$[-10, -0.5], and another with wider prior on \logWater{}: $\mathcal{U}$[-30, -0.5]. As compared to the retrieval without inclusion of \ch{H2O}, we obtain $\Delta$log$Z$ = 0.7 (2$\sigma$) for $\mathcal{U}$[-10, -0.5] and $\Delta$log$Z$ = 0.2 (1.4$\sigma$) for $\mathcal{U}$[-30, -0.5] indicating negligible sensitivity of the model comparison to the prior of \ch{H2O} abundance.

We discuss the implications of our marginal constraint on the presence and abundance of \ch{H2O} in Section \ref{sec:discuss_marginal_evidence_of_H2O}.          

\begin{figure}
\centering
\includegraphics[width=0.5\textwidth]{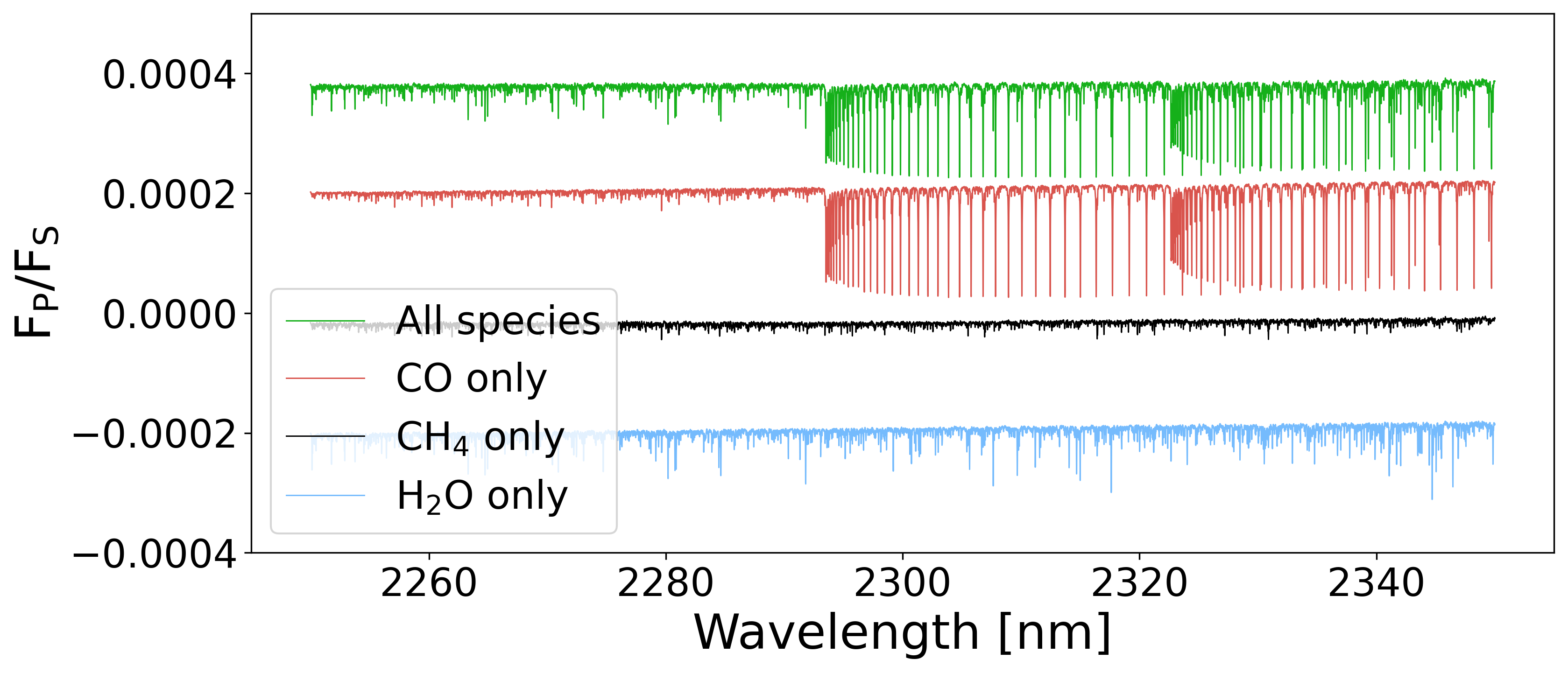}
\caption{Forward models of the emission spectrum of \taubootisb{} in the wavelength range of CRIRES observations analysed in this work. The models have been computed using {\tt GENESIS} corresponding to the best fit parameters (corresponding to the maximum a posteriori (MAP) solution in Table \ref{tab:param_priors_best_fit}) obtained from the atmospheric retrieval listed in Table \ref{tab:param_priors_best_fit} when accounting for all species (green) and for individual species (\ch{CH4} in black, \ch{CO} in red, and \ch{H2O} in blue), each shifted vertically for clarity.
}
\label{fig:forward_model}
\end{figure}

We compute the best fit forward model for the emission spectrum of the planet using {\tt GENESIS} by fixing the model parameters to their maximum a posteriori (MAP) values derived from the retrieval posteriors are shown in Figure \ref{fig:forward_model}. This includes contribution from individual species including \ch{CO}, \ch{H2O}, and \ch{CH4}. It is clear that the main contribution to the retrieved planetary model is from \ch{CO}, with a minor additional contribution from \ch{H2O} and no contribution from \ch{CH4}.  

\subsection{\KpVsys{} maps}
\label{sec:KpVsys_maps}
\begin{figure}
\centering
\includegraphics[width=0.5\textwidth]{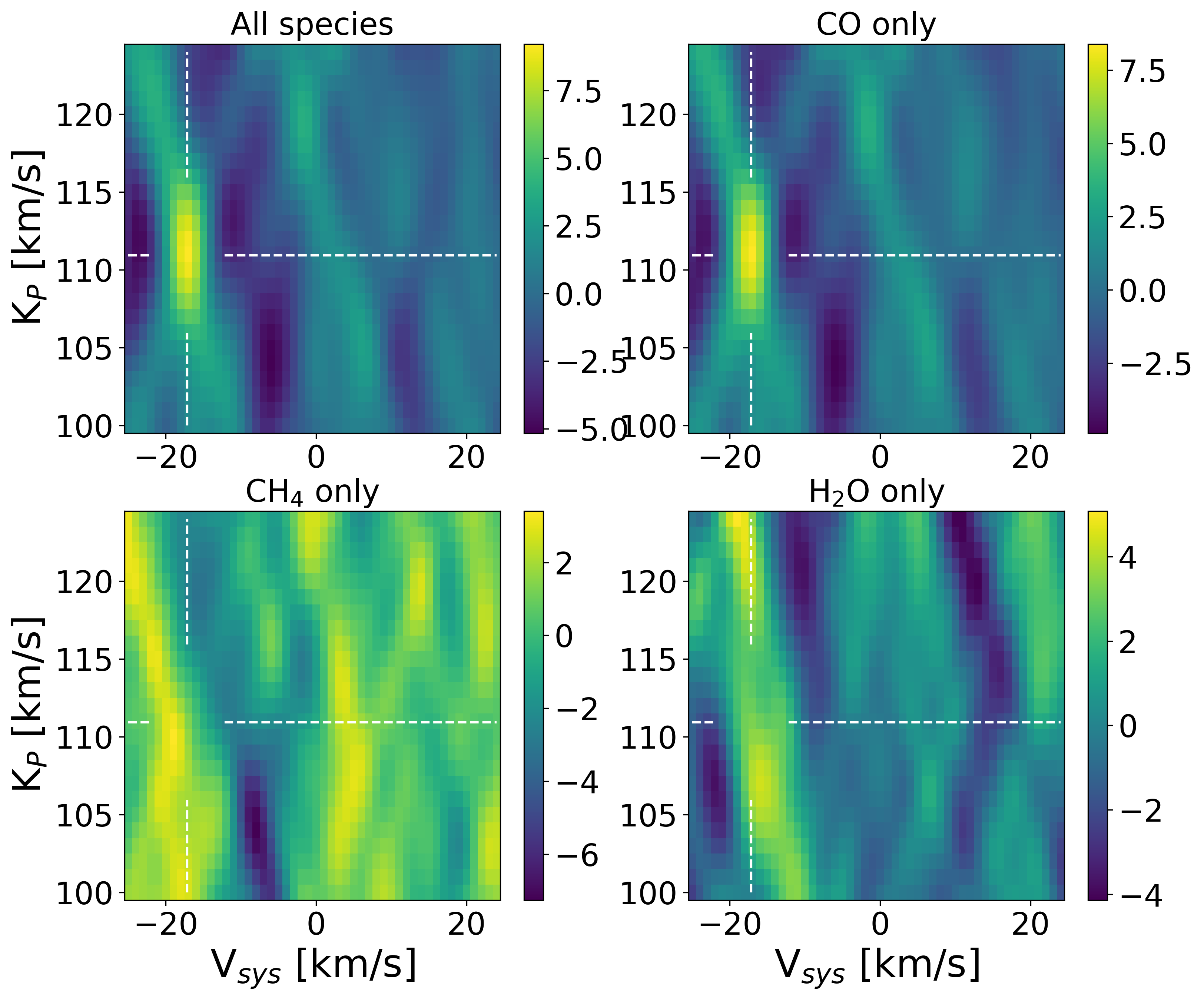}
\caption{Cross-correlation \KpVsys{} maps from all three nights of data combined, computed using the {\tt GENESIS} model (corresponding to the maximum a posteriori (MAP) solution in Table \ref{tab:param_priors_best_fit}). The colour bar shows the amplitude of the cross-correlation values. Going clockwise, the maps are computed using all species in the model, only CO, only \ch{CH4}, and only \ch{H2O}. The white horizontal and vertical lines mark the MAP \Kp{} and \Vsys{} values obtained from their marginalized posteriors. CO is clearly detected in the map and is the major driver of the detection, while the other two species do not show a clear signal at the expected \KpVsys{} location in the map.}
\label{fig:CC_maps}
\end{figure}

Another way to assess the detection significance of the atmospheric signal from \taubootisb{} is to compute CCF maps over a grid of \Kp{} and \Vsys{} values as described in Section \ref{sec:cross_correlation_logL}. In particular, we are interested in assessing the signal from all molecules in total, and from each of them individually. To do so, we first compute forward models for four cases as shown in Figure \ref{fig:forward_model}: 1) including all species, 2) including only \ch{CO}, 3) including only \ch{CH4}, 4) including only \ch{H2O}, with all model parameters fixed to their MAP values derived from the retrieval. Following the steps in Section \ref{sec:cross_correlation_logL} to calculate the CCF and the co-added across all exposures, detectors, and dates, we compute the CCF values over a \KpVsys{} grid with \Kp{} ranging from 90 to 140 \kms{}, and \Vsys{} ranging from -30 to 30 \kms{} in steps of 1 \kms{}. We repeat this for all the four cases. The resultant \KpVsys{} maps for the absolute CCF values are shown in Figure \ref{fig:CC_maps}. 

When including all species, there is a clear peak visible at the (\Kp{}, \Vsys{}) values derived from the retrieval. This looks indistinguishable from the \ch{CO} only case which indicates that \ch{CO} is the dominant source of the signal as expected. \ch{CH4} only case doesn't show any signal, whereas \ch{H2O} only case shows a smeared signal at low amplitude close to the expected \KpVsys{} value.

\begin{figure}
\centering
\includegraphics[width=0.5\textwidth]{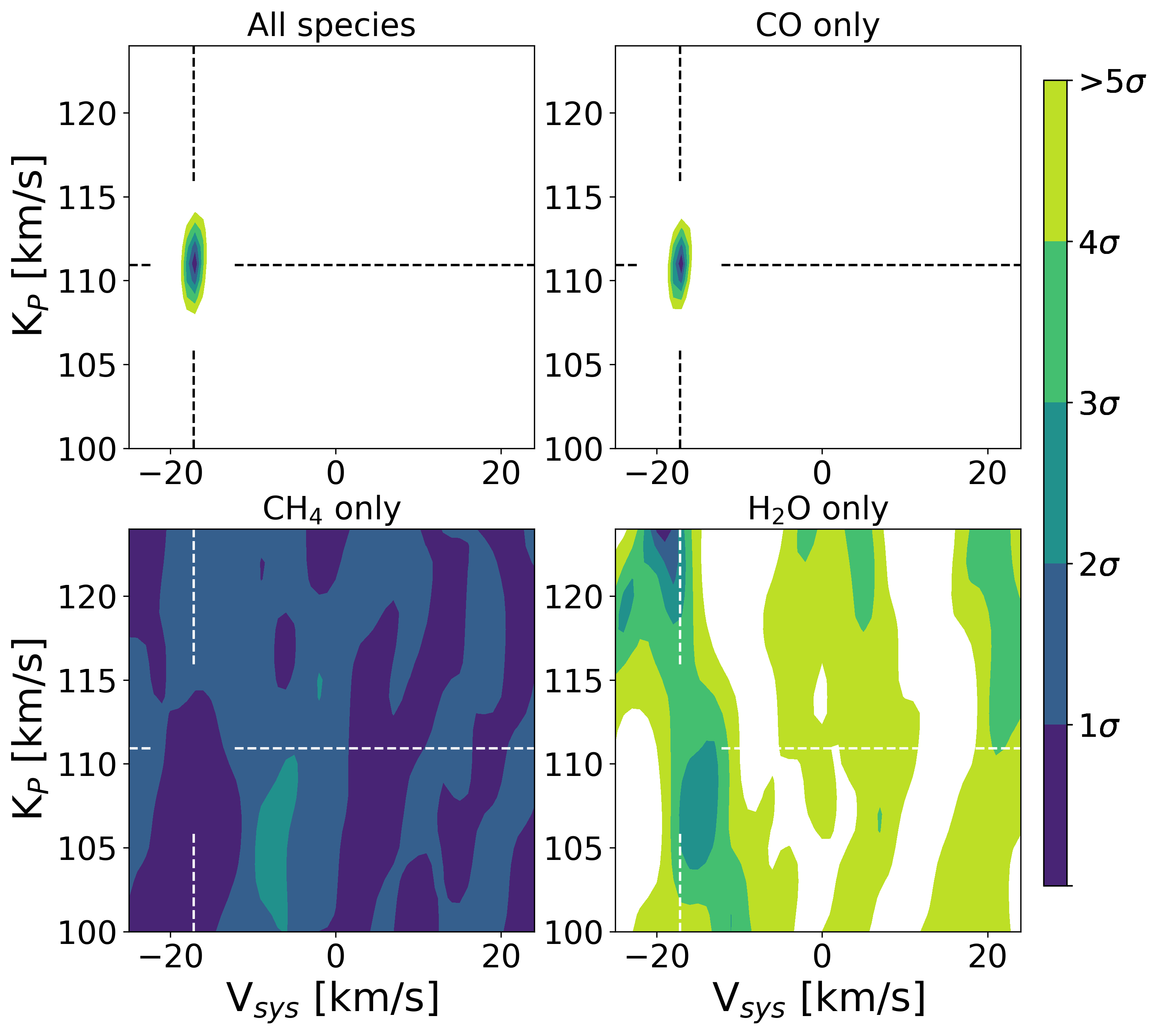}
\caption{CC-log (L) significance or confidence interval \KpVsys maps from all three nights of data combined, computed using {\tt GENESIS} model (corresponding to the maximum a posteriori (MAP) solution in Table \ref{tab:param_priors_best_fit}). Going clockwise, the maps are computed using all species in the model, only CO, only \ch{CH4}, and only \ch{H2O}. CO is detected at very high confidence ($\geq 5\sigma$), whereas there is no significant evidence for the other two species.}
\label{fig:logL_sigma_maps}
\end{figure}

To assess the statistical significance of the signal in the \KpVsys{} maps, we follow the \CCtologL{} map from Equation \ref{eq:cc2logL} to first construct a \KpVsys{} map of log-likelihood for all four models. We then use Wilks' theorem \citep{wilks_large-sample_1938} to calculate the chi-squared distribution for each point in the \KpVsys{} map as $\chi^{2}$ = --2$\Delta$log(L) = --2(log(L) -- log(L$_{max}$)). Here, log(L) is the log-likelihood at each point in the \KpVsys{} map and log(L$_{max}$) is the maximum likelihood across the map. We then compute the $p$ values relative to log(L$_{max}$) using the two tail survival function of the $\chi^{2}$ distribution with two degrees of freedom (for \Kp{} and \Vsys{}). We translate the $p$ values to confidence intervals or $\sigma$ level contours by computing the inverse survival function for normal distribution. The $\sigma$ level contours thus obtained are shown in Figure \ref{fig:logL_sigma_maps}. We emphasize that this method does not strictly yield a detection significance. We rather use confidence intervals between the peak log-likelihood value and the values around the peak in the rest of the map as a proxy for detection significance. In the likelihood framework, we can only assess how likely one model is relative to another model.   

The confidence interval maps show the atmospheric signal, the majority of which is from \ch{CO}, is detected at high significance ($>5\sigma$) at the retrieved \KpVsys{} position. Non-detection of \ch{CH4} is confirmed from the $1\sigma$ contour spread across the whole map. \ch{H2O} map shows a smeared 3$\sigma$ contour at the expected \KpVsys{} position, and intriguingly also shows a 1$\sigma$ peak offset in \Kp{} around 125 \kms{}, and another 2$\sigma$ contour around \Kp{} = 105 \kms{}, \Vsys{} = -16 \kms{}. 

\subsection{Spurious detection of \ch{CH4} due to imperfect telluric correction}
\label{sec:spurious_CH4}
\begin{figure}
\centering
\includegraphics[width=0.45\textwidth]{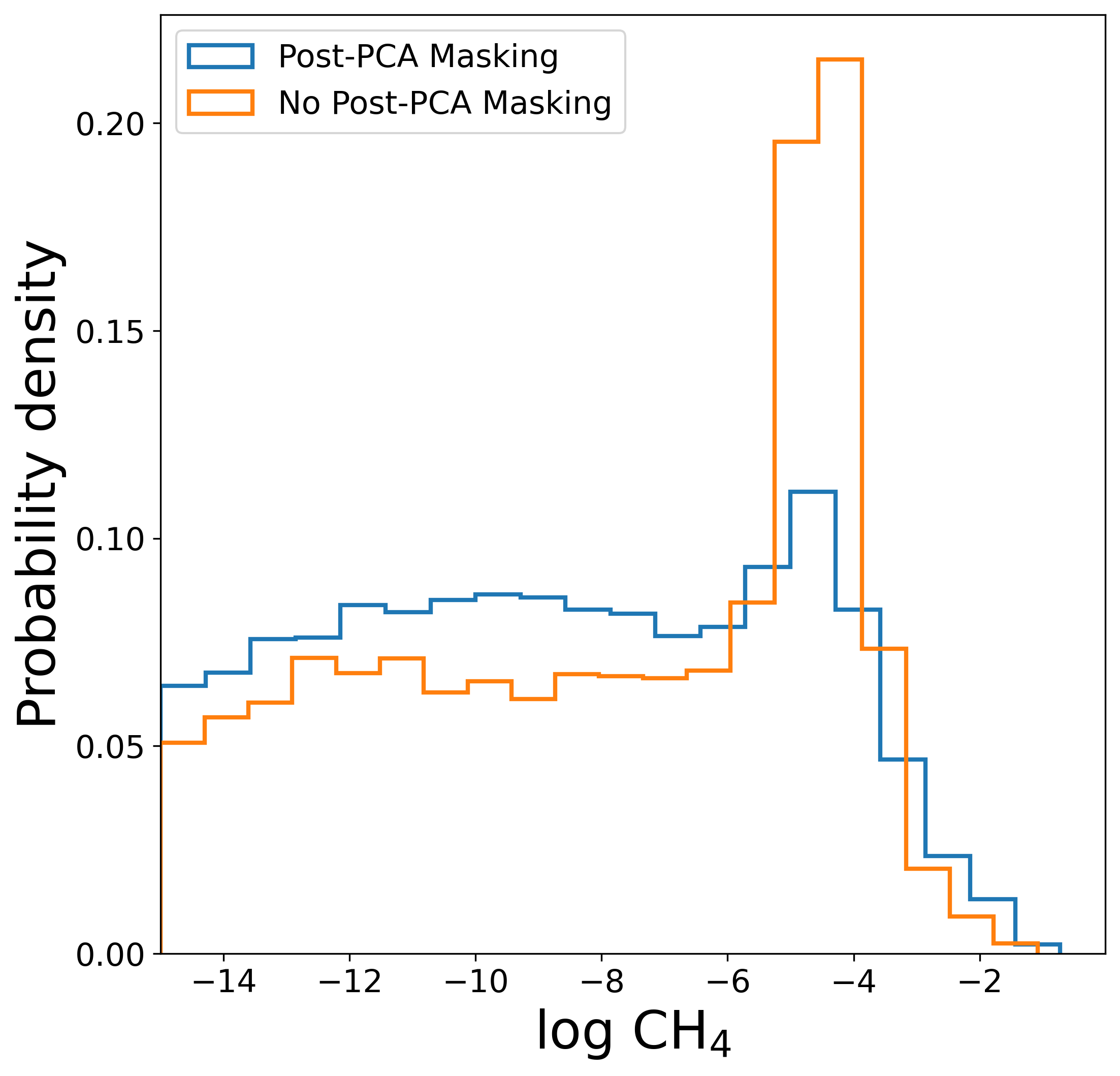}
\caption{Spurious peak in the posterior of methane abundance when not masking out noisy spectral channels after PCA detrending. These 1D posteriors are obtained from two retrieval runs, one which includes the post-PCA masking (in blue) and another which does not include the post-PCA masking (in orange).}
\label{fig:methane_1D_posteriors}
\end{figure}

An important step in removing telluric contamination in the data is masking spectral channels with strong telluric residuals remaining after PCA detrending as described in Section \ref{sec:post_PCA_mask}. We find that not masking after PCA detrending produces a peak in retrieval posteriors for \ch{CH4} abundance, which is likely due to telluric methane. In Figure \ref{fig:methane_1D_posteriors}, we show comparison between the marginalized posteriors for \ch{CH4} abundance when we do the post-PCA masking of noisy spectral channels, and when we do not. This shows that masking out insufficiently corrected spectral channels with significant level of residual telluric contamination is essential to avoid spurious detection of species common in both telluric and planetary spectrum.

\subsection{Inferred C/O and metallicity from retrievals}
\label{sec:C/O_metallicity}
Atmospheric C/O ratio and metallicity (referred to as [M/H]) are two key parameters that are believed to bear signatures of how a hot Jupiter like \taubootisb{} could have formed and migrated to its current position \citep{madhusudhan_toward_2014}. The posterior distributions for the molecular species included in the retrieval analysis can be converted to posterior distributions for atmospheric C/O ratio and [M/H] through stoichiometry. To compute the C/O ratio, for each sample point in the {\tt pymultinest} posterior samples we compute C/O ratio as:  

\begin{equation}\label{eq:C_to_O}
\mathrm{C/O} = \frac{[\ch{CO}] + [\ch{CH4}] + [\ch{CO2}] + [\ch{HCN}]}{ [\ch{CO}] + [\ch{H2O}] + 2[\ch{CO2}] }
\end{equation}

where numerator is the VMR of Carbon atoms, and denominator is the VMR of Oxygen atoms. 

To compute [M/H], for each sample point in the posterior, we first compute the VMR of metals as:

\begin{equation}\label{eq:metals_VMR}
[\mathrm{M}] = 2[\ch{CO}] + [\ch{H2O}] + [\ch{CH4}] + 3[\ch{CO2}] + 2[\ch{HCN}] + [\ch{NH3}]
\end{equation}

and number density of Hydrogen atoms as 
\begin{equation}\label{eq:hydrogen_VMR}
[\mathrm{H}] = 2 \frac{(1 -  [\ch{CO}] + [\ch{H2O}] + [\ch{CH4}] + [\ch{CO2}] + [\ch{HCN}] + [\ch{NH3}])}{1.176}
\end{equation}

where the numerator is the VMR of the \ch{H2}-\ch{He} mixture, and the denominator accounts for the fraction of which is only \ch{H2} gas. The factor 2 is to account for two Hydrogen atoms per molecule of Hydrogen gas. [M/H] is then calculated as log the ratio of [M] to [H]. 

\begin{figure}
\centering
\includegraphics[width=0.5\textwidth]{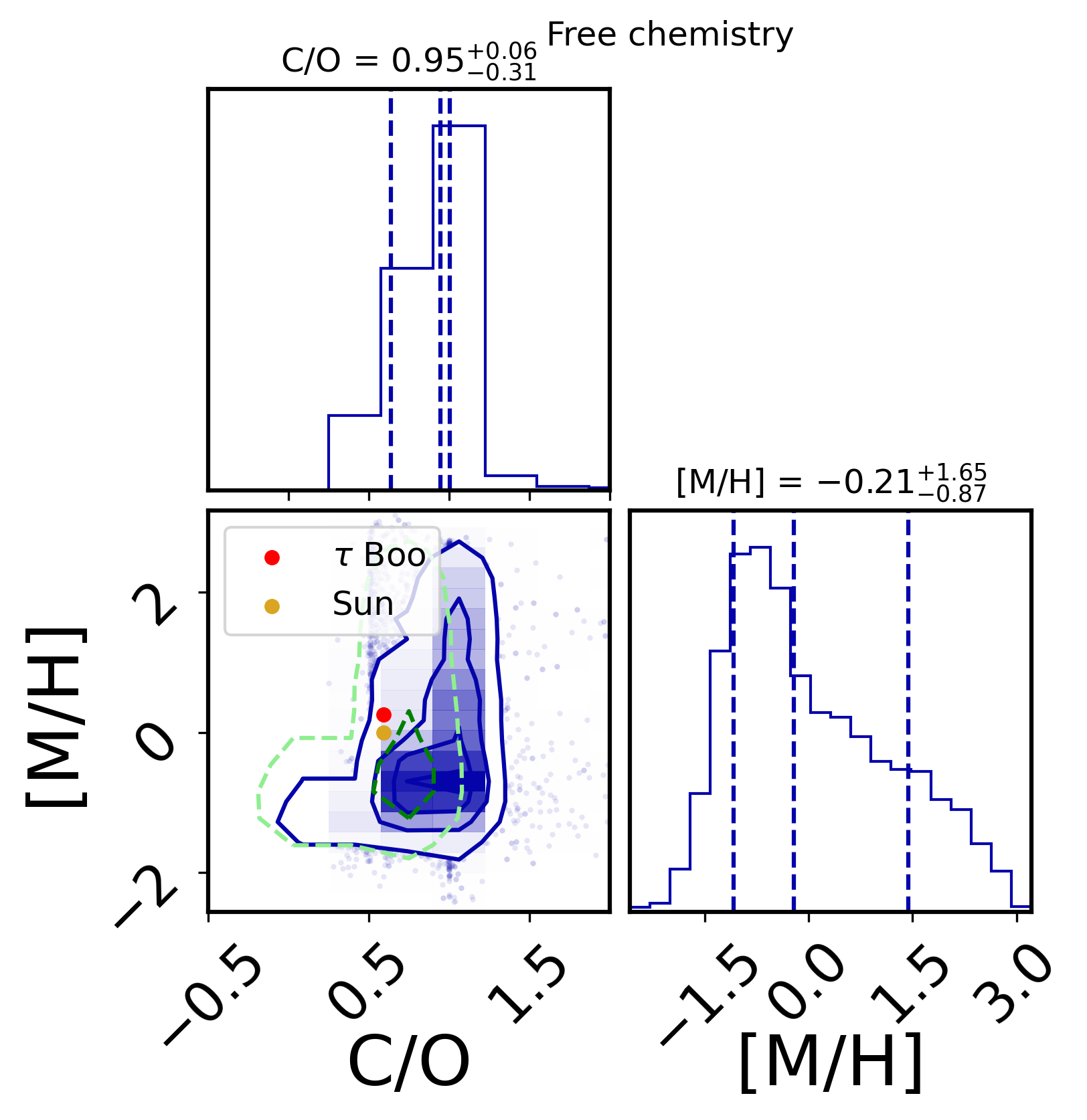}
\caption{Corner plots showing the 1D and 2D posterior distributions for the C/O ratio and metallicity ([M/H] in dex units) which we obtain from the posterior distributions of each molecular species included in the free chemistry retrieval in this work. The yellow dot marks the C/O and [M/H] of the Sun \citep{asplund_chemical_2021} and the red dot is the same for \taubootisStar{} (stellar metallicity from \protect \cite{borsa_gaps_2015}, stellar C/O ratio assumed to be solar). In the 2D posterior subplot, the blue contours mark the $1\sigma$, $2\sigma$, and $3\sigma$ levels of the distribution, while the dashed green contours show the same for when accounting for 20\% Oxygen sequestered due to condensation in silicate clouds.}

\label{fig:co_met_posterior}
\end{figure}

Since at the typical temperatures of tau Boo the atmosphere crosses the condensation curves of several minerals incorporating oxygen (e.g. \citep{burrows_chemical_1999, wakeford_high-temperature_2017}), the measured abundance of Oxygen in the gas phase might underestimate the true Oxygen abundance. Hence, we repeat the calculation, accounting for additional 20\% Oxygen that may be sequestered in silicate clouds. The resulting 2D and 1D posteriors for C/O ratio and [M/H] are shown in Figure \ref{fig:co_met_posterior} along with comparison to C/O ratio and [M/H] of the Sun and \taubootisStar{}. From these posteriors, we infer a super solar C/O = 0.95$^{+0.06}_{-0.31}$ at high confidence \citep[solar C/O = 0.59 $\pm$ 0.08, ][]{asplund_chemical_2021}, and sub-solar metallicity [M/H] = -0.21 $^{+1.66}_{-0.87}$ dex. Accounting for 20$\%$ Oxygen sequestration in silicate clouds changes these estimates slightly within 1$\sigma$ uncertainties : C/O = 0.79$^{+0.05}_{-0.26}$ and [M/H] = -0.2 $^{+1.66}_{-0.86}$. 

\begin{figure}
\centering
\includegraphics[width=0.5\textwidth]{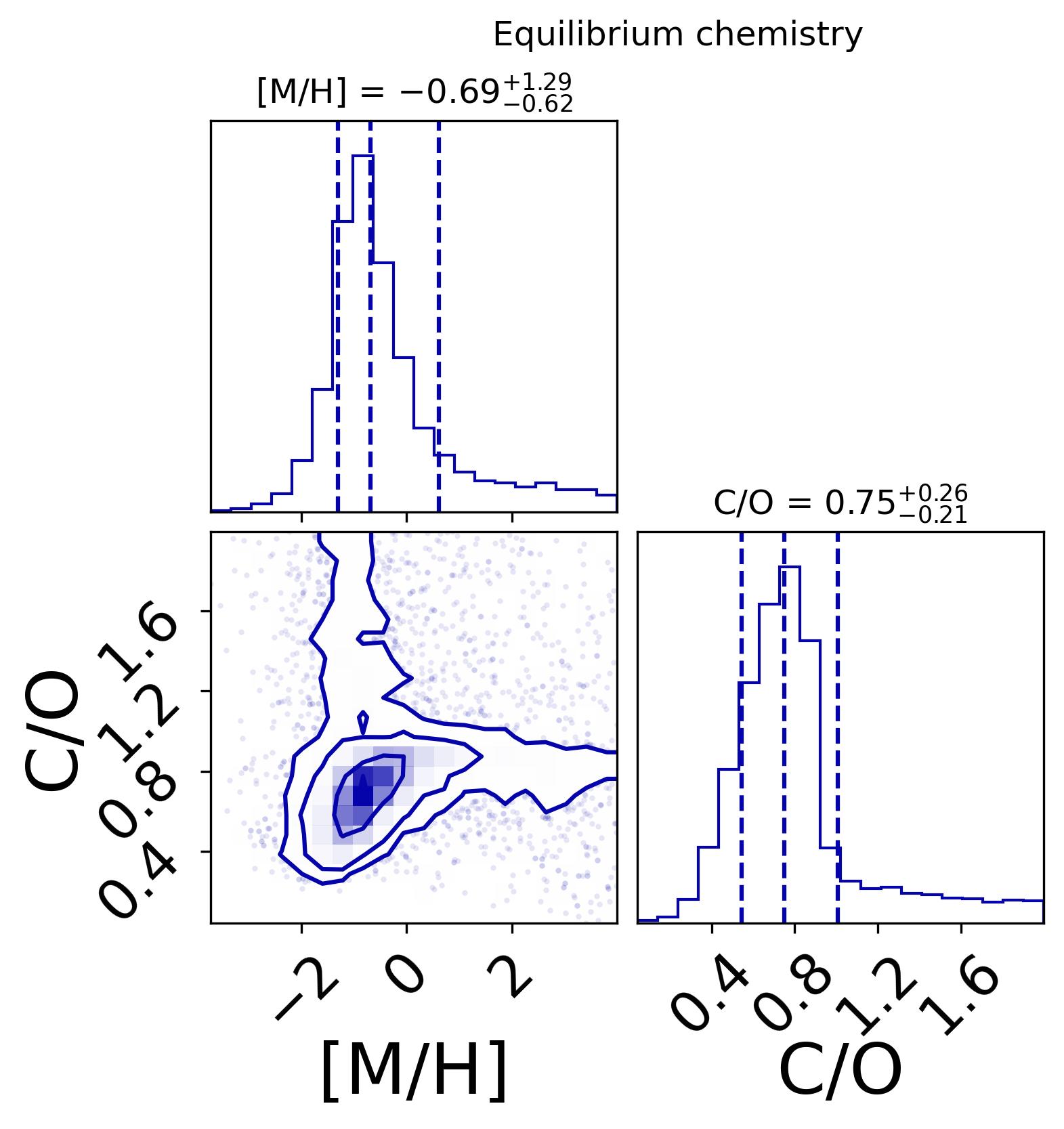}
\caption{Corner plots showing the 1D and 2D posterior distributions for the C/O ratio and metallicity ([M/H] in dex units) obtained from equilibrium chemistry retrieval.}

\label{fig:co_met_posterior_eq}
\end{figure}

To confirm that the super-solar C/O is not biased by the strong detection of CO and weak detection of \ch{H2O}, we also run an equilibrium chemistry retrieval by fitting directly for C/O and metallicity. The computational steps for equilibrium chemistry retrieval are similar to the free chemistry retrieval except how the abundances of individual molecules used for computing the emission spectrum. For performing the equilibrium chemistry retrieval, we use \texttt{FastChem} \citep{stock_fastchem_2018, stock_fastchem_2022} to compute the equilibrium chemistry abundance profiles for each molecule for given C/O, metallicity, and \TP{} profile, and input them to \texttt{GENESIS} to compute the emission spectrum corresponding to equilibrium chemistry. When setting up \texttt{FastChem}, for a given metallicity, we first scale the abundance of all the elements, and then adjust the abundance of Carbon with respect to the metallicity adjusted Oxygen abundance according to the specified value of C/O. The resultant posteriors for C/O and [M/H] obtained are shown in Figure \ref{fig:co_met_posterior_eq}. We find C/O = 0.75$^{+0.26}_{-0.21}$ and [M/H] = --0.69 $^{+1.29}_{-0.62}$, consistent with the respective values obtained from free-chemistry retrieval within 1$\sigma$.

\section{Discussion}
\label{sec:discussion}
\subsection{Confirmation of CO}
\label{sec:discuss_confirmation_of_CO}
Using three nights of data spanning the wavelength range of 2287 to 2345 nm in K band, we detect a significant signal from the atmosphere of \taubootisb{}, stemming largely from \ch{CO} absorption lines. This confirms the previous detection of \ch{CO} from the same dataset first by \cite{brogi_signature_2012} and is consistent with the detection by \cite{pelletier_where_2021} from SPIRou observations. The CO abundance constrained by our retrievals of \logCO{} = --3.44$^{+1.63}_{-0.85}$ is lower than the slightly super-solar value of --2.46$^{+0.25}_{-0.29}$ recovered by \citep{pelletier_where_2021}, but is still consistent with it within 1$\sigma$. We argue that the relatively lower precision on our retrieved value of CO abundance as compared to \cite{pelletier_where_2021} despite having very high SNR observations is primarily because of the narrow wavelength range covered by CRIRES data as compared to SPIRou. Additionally, focussing on wavelength coverage of dominant CO lines alone, SPIRou does not gain much in terms of spectral range compared to the old CRIRES. Furthermore, the CFHT is half the size of the VLT, which means the SPIRou observations are at half the S/N as compared to CRIRES. However, with \taubootisStar{} being a northern target, the VLT AO system MACAO underperforms when being forced to operate at relatively high airmass. SPIRou has a high efficiency by design, which could also be a contributing factor.  

\subsection{Marginal evidence for \ch{H2O}}
\label{sec:discuss_marginal_evidence_of_H2O}
Based on the tentative peak observed in the retrieval posteriors for \ch{H2O} abundance, we find a marginal evidence for water in the atmosphere of \taubootisb{}. However, given that by using the model for the best fit \logWater{} = --4.67 VMR we do not obtain a detection in the \KpVsys{} maps (Figure \ref{fig:CC_maps} and \ref{fig:logL_sigma_maps}), it could mean that our inferred posteriors for \ch{H2O} abundance and its 3$\sigma$ upper limit needs further verification from observations that span a larger wavelength coverage, especially in the J and H where \ch{H2O} lines are most dominantly present as compared to \ch{CO}. Further observations can also test the peak in the posterior and confirm or rule out potential contamination from residual telluric \ch{H2O} lines in the data.

The 3$\sigma$ upper limit constrained from our retrieval at \logWater{} = --1.12 is higher than the value \logWater{} = --5.66 obtained by \cite{pelletier_where_2021}. However, the tail of the 1D posterior for \ch{H2O} abundance from our retrieval towards lower abundances means that our posterior is consistent with the SPIRou measurement. Due to the significant tail towards lower abundances, we do not quote the lower 1$\sigma$ limit for \ch{H2O} as it would depend on the lower limit on the prior. For the same reason, it is not possible to translate the \ch{H2O} abundance posterior into a lower limit. For the 3 sigma upper limit we quote, however, this effect would be minor. In summary, despite the fact that we find a marginal peak for \logWater{} at relatively high VMRs compared to the measurements made by SPIRou, the two measurements are fully consistent.

Our marginal evidence for \ch{H2O} aligns with the detection of \ch{H2O} in the CARMENES data by \cite{webb_water_2022} who detect a signal in the CCF maps for a model with \logWater{} = --3. A note of caution for the reader here is that \cite{webb_water_2022} do not perform a retrieval and therefore their quoted VMR is for a fixed \TP{} profile chosen ad hoc. Due to the degeneracy between \TP{} profile and VMRs, their quoted \logWater{} = --3 can not be taken at face value for comparison with our measurements. 

\subsection{Super-solar C/O and slightly sub-solar [M/H] of \taubootisb{}}
\label{sec:discuss_high_CO_met}
Our retrievals constrain a super solar C/O ratio = 0.95$^{+0.06}_{-0.31}$, and sub-solar median [M/H] = --0.21 $^{+1.66}_{-0.87}$ (0.62$^{+2.37}_{-1.24}$ $\times$ solar) for the atmosphere of \taubootisb{}. This median [M/H] value is lower than the super-solar values of [C/H] (5.85 $^{+4.44}_{-2.82}$) and [O/H] (3.21 $^{+2.43}_{-1.56}$) constrained by \cite{pelletier_where_2021} and is driven by the higher upper limit and the tentative peak in the \logWater{} constrained by our retrievals. However, considering the large $1\sigma$ uncertainty on our measurement, we cannot rule out the super-solar metallicity found by \cite{pelletier_where_2021} at high confidence. We highlight here that, similar to this work, \cite{pelletier_where_2021} also assumed well-mixed chemistry and abundances of all the molecules in their models to be constant with pressure. 

In a simple picture of standard core-accretion model of formation for gas-giants \citep{pollack_formation_1996}, a giant planet is expected to undergo either gas-dominated accretion or planetesimal dominated accretion. Gas dominated accretion beyond the water ice line can end up in a high C/O, low [M/H] planet, reflecting the typical C/O distribution of gas in the disk \citep[e.g.][]{oberg_effects_2011}. On the other hand, planetesimal dominated accretion with delivery of metal rich solids (as the planet contracts and migrates through the disk) is expected to result in high [M/H] but low C/O ratio planet \citep[e.g.][]{madhusudhan_toward_2014, khorshid_simab_2022}. 

The super-solar C/O ratio and [M/H] for \taubootisb{} inferred from SPIRou observations by \cite{pelletier_where_2021} is incompatible with this picture, and requires additional mechanisms e.g. pebble drift, disk evolution, fractionation of accreted material between the planetary core and the envelope to name a few that can enrich the \taubootisb{} metallicity. In comparison, the super-solar C/O ratio and median sub-solar [M/H] for \taubootisb{} observed in our work seems to align well with the gas-dominated case of core-accretion. The caveat here is the large 1$\sigma$ uncertainty on the [M/H] which does not confidently rule out the scenario consistent with SPIRou observations.         

\subsection{Potential contamination from the night side}
Since \taubootisb{} is a non-transiting planet with inclination of 43.92$\pm$0.52 $^{\circ}$, the geometric configuration of the system means that around the phase 0.5 when our observations are taken, the contribution from the night side of the planet is not negligible. We use {\tt starry} \citep{luger_starry_2019} to compute a broadband phase curve of \taubootisb{} and find that at phase = 0.5, there is a $\sim$30 \% dip in peak in planetary flux relative to the hypothetical case if \taubootisb{} had an inclination of 90 $^{\circ}$. This is due to $\sim$30 \% dip in flux at phase 0.5 arising from the fraction of the nightside visible. This implies that there is potential contribution from the night side in our observations, and using a 1D model with a single \TP{} profile may not be sufficient. Simulations by  \cite{beltz_significant_2021} for CRIRES observations of HD209458 b show that \TP{} the profile for a typical hot-Jupiter can vary significantly across the longitudes, showing 500-700 K maximum variation in temperature around 10$^{-2}$ bars. Not accounting for this multidimensionality in \TP{} profiles could influence the detection significances \ch{H2O} and \ch{CO}. We recommend further analysis for this and future data sets of a non-transiting planet like \taubootisb{} to account for the longitudinal variation of the \TP{} profile when computing the model for the emission spectrum used for calculating the CCFs. 

\section{Conclusion}
In this work, we set out to resolve the tension about the presence of water \ch{H2O} in the atmosphere of the hot Jupiter \taubootisb{}. We reanalysed the archival CRIRES observations of \taubootisb{} using the currently prevalent techniques. This includes detrending the telluric and instrumental systematics using PCA, and constraining the abundances of molecules, including \ch{H2O} and \ch{CO} using cross-correlation to log-likelihood mapping and Bayesian retrievals. 

Based on our analyses in this work, our main conclusions are: 
\begin{itemize}
  \item We introduce a new metric to optimize the choice of number of PCA components for detrending HRCCS data. This metric ensures the choice of PCA components leads to optimal suppression of telluric noise in the cross-correlation, rather than maximum recovery of planetary signal, as the latter can also amplify the local noise along with optimizing the recovery of the planetary signal.     
  \item We confirm detection of CO, and non-detection of \ch{H2O} and \ch{CH4}. From our retrieval analysis, we obtain constraints on \Kp{} and \Vsys{} for the system consistent with previous observations. 
  \item We obtain a tentative peak in the retrieval posteriors for \ch{H2O} abundance, with a 3$\sigma$ upper limit on the \logWater{} = --1.12. Our posterior for \logWater{} overlap with the 3$\sigma$ upper limit reported by \cite{pelletier_where_2021} (\logWater{} = --5.66) and our results are consistent with \cite{pelletier_where_2021} within 1$\sigma$.
  \item From free chemistry retrieval, we infer super-solar C/O (0.95$^{+0.06}_{-0.31}$), marginally sub-solar [M/H] (--0.21 $^{+1.66}_{-0.87}$) albeit with large 1$\sigma$ uncertainty that doesn't rule out the previously reported super-solar metallicity by \citep{pelletier_where_2021}. We also perform an equilibrium chemistry retrieval and find values for C/O and [M/H] consistent within 1$\sigma$ of those obtained from free chemistry retrieval.
  \item We find that insufficient telluric correction can lead to spurious peaks in the retrieval posteriors for a molecule like \ch{CH4} which is common to both the planetary atmosphere and earth's atmosphere. We conclude that masking out insufficiently detrended spectral channels after performing PCA masking can help to avoid this. 
\end{itemize}

While we confirm the super-solar C/O, the large uncertainty on metallicity prevents us from confirming or refuting conclusively whether the planet has super-solar metallicity that goes against standard core-accretion, or sub-solar metallicity more in line with expectations from core accretion. The main cause of limitation in this work is the narrow spectral wavelength range of the old CRIRES data, where we have contribution only from the water lines. This can be overcome by the analysis of new CRIRES+ observations, which cover a much larger wavelength range simultaneously. 

We also speculate that the non-transiting geometry of the planet leads to a non-negligible contribution from the night side. This implies that analysis using models for only the day side of the planet may not be sufficient in capturing the complex global thermal structure of the planet contributing to the visible emission spectrum. Accounting for contributions from the night side using sophisticated 3D modeling and retrievals could be beneficial in this context.

Brightness of \taubootisStar{} makes it amenable to observations in the M band using CRIRES+ where features from refractory content through SiO lines can be probed, and the planet's Si/H and Si/O ratios can be measured. Simultaneous constraint on the Si/H and C/O ratio from the such observations can shed further light on the super-solar C/O of the planet, and test the hypothesis of whether measured super-solar C/O for tau Boo could be a result of sequestration of oxygen in silicates, which is an effect more pronounced at higher Si/H \citep{chachan_breaking_2023}. 

\section*{Acknowledgements}
We thank the anonymous reviewer for their feedback and comments which greatly helped in improving the clarity and thoroughness of this work. We thank Jayne Birkby, Lennart van Sluijs, Luke Parker, and Spandan Dash for insightful discussions on the analysis and interpretation of CRIRES data. V.P. acknowledges support from the UKRI Future Leaders Fellowship grant MR/S035214/1 and UKRI Science and Technology Facilities Council (STFC) through the consolidated grant ST/X001121/1. This research was in part funded by the UKRI Grant EP/X027562/1.
\section*{Data Availability}
The raw science data included in this work is openly available on the ESO Science archive. The processed data, analysis outputs, and the code used to make the figures in the paper, are available on Zenodo\footnote{10.5281/zenodo.11570451}. 




\bibliographystyle{mnras}
\bibliography{refs} 




\appendix
\section{Selecting the optimal number of PCA components}
\begin{figure*}
\centering
\includegraphics[width=1\textwidth]{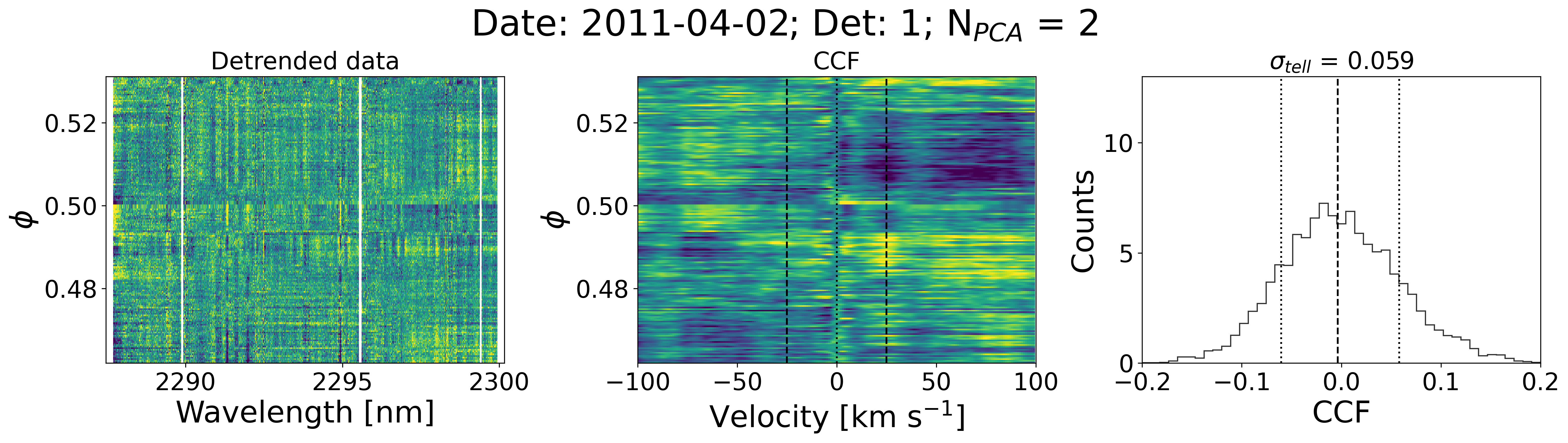}
\includegraphics[width=1\textwidth]{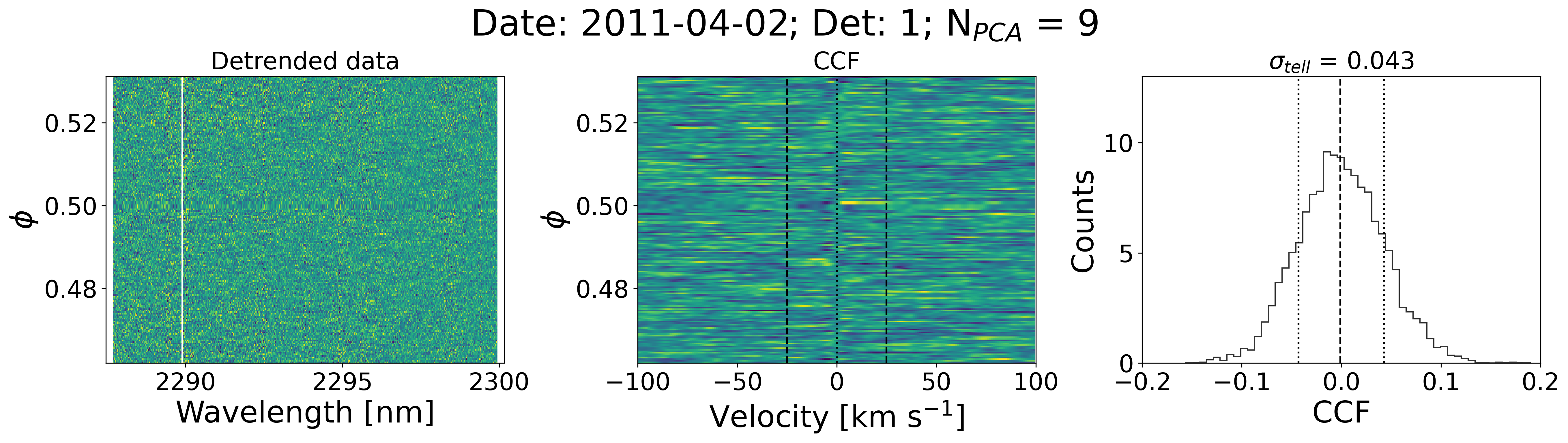}
\caption{Same as Figure \ref{fig:N_PCA_optimization_demo} but for two values of \Npca{} (2 for the top three sub-figures, and 9 for the bottom three sub-figures). In both top and bottom sub-figures, we show the PCA detrended data cube (left panel), CCF of the data with an ESO SkyCalc model (middle panel) for each phase, and the distribution of the CCF values in the $\pm$25 \kms{} region around 0 \kms{} (right panel). The width of the CCF distribution \sigmatell{} decreases as \Npca{} increases from 2 to 9, indicating improvement in removal of telluric lines with increasing \Npca{}.}
\label{fig:N_PCA_optim_trail_matrix_histogram}
\end{figure*}

In Section \ref{sec:compute_optimal_PCA} we describe our steps of computing the optimal \Npca{} which provides sufficient suppression of telluric noise in the data. Similar to Figure \ref{fig:N_PCA_optimization_demo}, in this section we show the steps towards the derivation of our metric \sigmatell{} for two values of \Npca{} (see Figure \ref{fig:N_PCA_optim_trail_matrix_histogram}). \Npca{} = 2 represents the minimum level of PCA detrending which doesn't remove telluric contamination in the data and yields a relatively high value of \sigmatell{}. Increasing \Npca{} to 9, we observe that the telluric contribution in the CCF diminishes and leads to relatively smaller value of \sigmatell{}. 

\section{Full posterior distribution from retrievals}
\begin{figure*}
\centering
\includegraphics[width=1.05\textwidth]{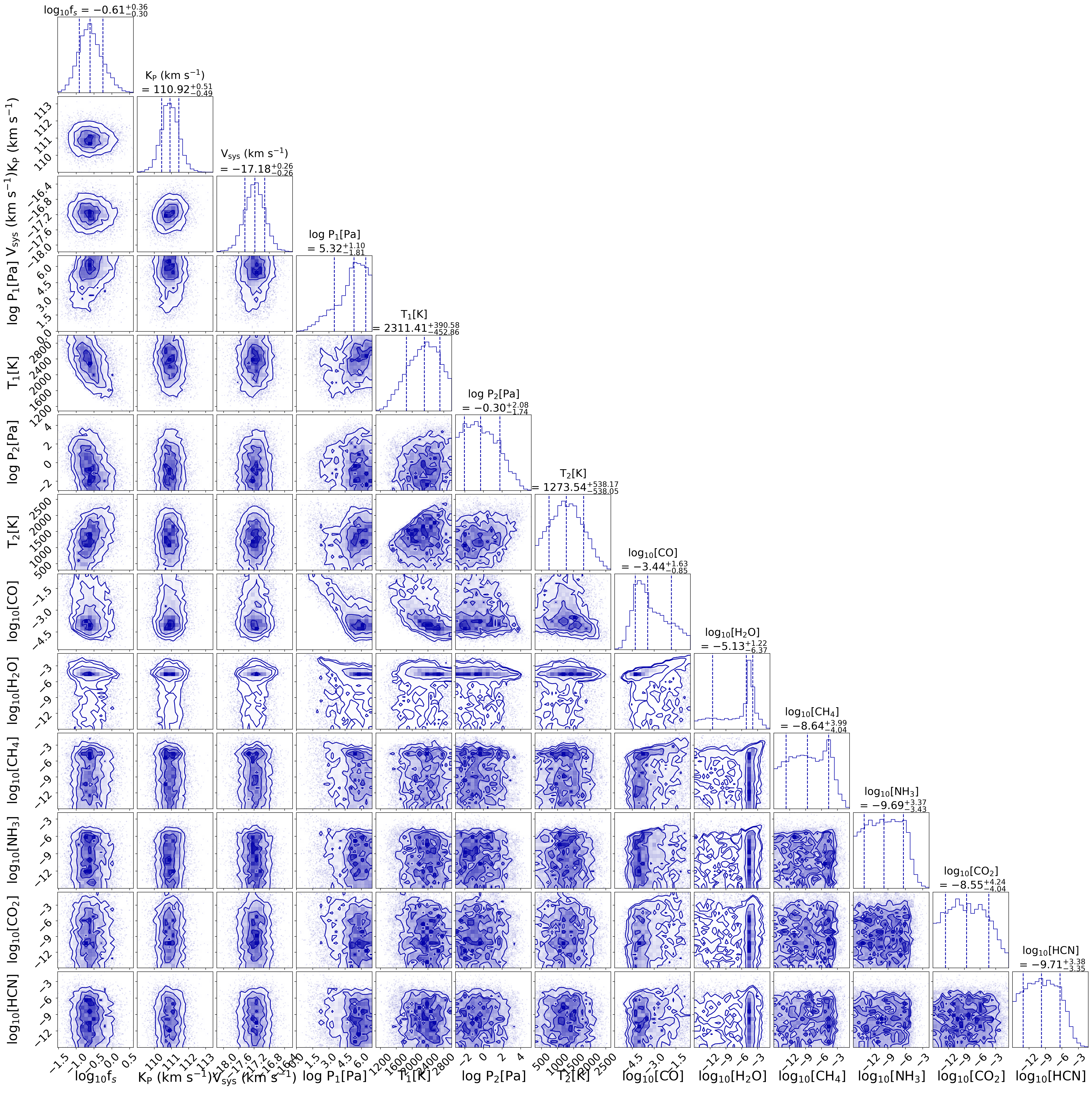}
\caption{Full posterior distribution showing all the free parameters in the joint retrieval analyses of all three nights of data for \taubootisb{}. The median and $\pm1\sigma$ values, and the maximum a posteriori (MAP) solution of each parameter obtained from this retrieval are labelled here and also shown in Table \ref{tab:param_priors_best_fit}.  The zoomed in posteriors of \Kp{} and \Vsys{} are shown in Figure \ref{fig:KpVsys_posteriors}, while those for model scaling factor and abundances of selected molecular species are shown in Figure \ref{fig:abundance_posteriors}.}
\label{fig:retrieval_posteriors_full}
\end{figure*}

In this section, we show the full posterior distribution obtained from our retrieval analysis described in Section \ref{sec:atmospheric_retrieval_setup} for analysis using \Npca{} = 9. We show the corner plots for the posterior distributions and correlation between all the free parameters in the retrieval in Figure \ref{fig:retrieval_posteriors_full}.


\bsp	
\label{lastpage}
\end{document}